\documentclass[iop]{emulateapj}
\usepackage{apjfonts,color}
\journalinfo{Accepted for publication in the Astrophysical Journal}
\slugcomment{Submitted to ApJ September 17, 2011; accepted July 9, 2012}

\newcommand{\erfc}{\ensuremath{\mathrm{erfc}}}

\newcommand{\Msol}{\ensuremath{M_\odot}}

\newcommand{\etal}{\emph{et al.}}
\newcommand{\aanda}{A\&A}
\newcommand{\astroph}[1]{arXiv:astro-ph/{#1}}
\newcommand{\arxiv}[1]{arXiv:{#1}}

\newcommand{\nickel}{{\ensuremath{^{56}\mathrm{Ni}}}}
\newcommand{\cobalt}{{\ensuremath{^{56}\mathrm{Co}}}}

\newcommand{\vSi}{\ensuremath{v}(\ion{Si}{2})}
\newcommand{\kms}{\ensuremath{\mathrm{km~s}^{-1}}}
\newcommand{\lcdm}{\ensuremath{\Lambda}CDM}

\begin{document}


\title{A Search for New Candidate Super-Chandrasekhar-Mass
       Type I\lowercase{a} Supernovae \\
       in the Nearby Supernova Factory Dataset}
      
\author
{
    R.~Scalzo,\altaffilmark{1}
    G.~Aldering,\altaffilmark{2}
    P.~Antilogus,\altaffilmark{3}
    C.~Aragon,\altaffilmark{2,4}
    S.~Bailey,\altaffilmark{2}
    C.~Baltay,\altaffilmark{5}
    S.~Bongard,\altaffilmark{3}
    C.~Buton,\altaffilmark{6}
    A.~Canto,\altaffilmark{3}
    F.~Cellier-Holzem,\altaffilmark{3}
    M.~Childress,\altaffilmark{2,7}
    N.~Chotard,\altaffilmark{8}
    Y.~Copin,\altaffilmark{8}
    H.~K. Fakhouri,\altaffilmark{2,7}
    E.~Gangler,\altaffilmark{8}
    J.~Guy,\altaffilmark{3} 
    E.~Y. Hsiao,\altaffilmark{2}
    M.~Kerschhaggl,\altaffilmark{6}
    M.~Kowalski,\altaffilmark{6}
    P.~Nugent,\altaffilmark{9}
    K.~Paech,\altaffilmark{6}
    R.~Pain,\altaffilmark{3}
    E.~Pecontal,\altaffilmark{10}
    R.~Pereira,\altaffilmark{8}
    S.~Perlmutter,\altaffilmark{2,7}
    D.~Rabinowitz,\altaffilmark{5}
    M.~Rigault,\altaffilmark{8}
    K.~Runge,\altaffilmark{2}
    G.~Smadja,\altaffilmark{8}
    C.~Tao,\altaffilmark{11,12}
    R.~C. Thomas,\altaffilmark{9}
    B.~A.~Weaver,\altaffilmark{13}
    \& C.~Wu\altaffilmark{3,14} \\
    (The~Nearby~Supernova~Factory)
}
\affil{}
\email{rscalzo@mso.anu.edu.au}

\altaffiltext{1}
{
    Research School of Astronomy and Astrophysics,
    The Australian National University,
    Mount Stromlo Observatory,
    Cotter Road, Weston Creek ACT 2611 Australia
}
\altaffiltext{2}
{
    Physics Division, Lawrence Berkeley National Laboratory, 
    1 Cyclotron Road, Berkeley, CA, 94720
}
\altaffiltext{3}
{
    Laboratoire de Physique Nucl\'eaire et des Hautes \'Energies,
    Universit\'e Pierre et Marie Curie Paris 6,
    Universit\'e Paris Diderot Paris 7, CNRS-IN2P3, 
    4 place Jussieu, 75252 Paris Cedex 05, France
}
\altaffiltext{4}
{
    Present address:  Department of Human Centered Design \& Engineering,
    University of Washington,
    423 Sieg Hall, Box 352315, Seattle, WA 98195
}
\altaffiltext{5}
{
    Department of Physics, Yale University, 
    New Haven, CT, 06250-8121
}
\altaffiltext{6}
{
    Physikalisches Institut, Universit\"at Bonn,
    Nu\ss allee 12, 53115 Bonn, Germany
}
\altaffiltext{7}
{
    Department of Physics, University of California Berkeley,
    366 LeConte Hall MC 7300, Berkeley, CA, 94720-7300
}
\altaffiltext{8}
{
    Universit\'e de Lyon, F-69622, Lyon, France;
    Universit\'e de Lyon 1, Villeurbanne; 
    CNRS/IN2P3, Institut de Physique Nucl\'eaire de Lyon.
}
\altaffiltext{9}
{
    Computational Cosmology Center, Computational Research Division,
    Lawrence Berkeley National Laboratory, 
    1 Cyclotron Road MS 50B-4206, Berkeley, CA, 94720
}
\altaffiltext{10}
{
    Centre de Recherche Astronomique de Lyon, Universit\'e Lyon 1,
    9 Avenue Charles Andr\'e, 69561 Saint Genis Laval Cedex, France
}
\altaffiltext{11}
{
    Centre de Physique des Particules de Marseille,
    163, avenue de Luminy - Case 902 - 13288 Marseille Cedex 09, France
}
\altaffiltext{12}
{
    Tsinghua Center for Astrophysics,
    Tsinghua University, Beijing 100084, China 
}
\altaffiltext{13}
{
    Present address:  Center for Cosmology and Particle Physics,
    New York University, 4 Washington Place, New York, NY 10003, USA
}
\altaffiltext{14}
{
    National Astronomical Observatories,
    Chinese Academy of Sciences, Beijing 100012, China
}

\shorttitle{SNfactory Observations of Candidate Super-Chandra SNe~Ia}
\shortauthors{Scalzo et al.}

\begin{abstract}
We present optical photometry and spectroscopy of five type Ia supernovae
discovered by the Nearby Supernova Factory selected to be spectroscopic
analogues of the candidate super-Chandrasekhar-mass events SN~2003fg and
SN~2007if.  Their spectra are characterized by hot, highly ionized
photospheres near maximum light, for which SN~1991T supplies the best phase
coverage among available close spectral templates.
Like SN~2007if, these supernovae are overluminous ($-19.5 < M_V < -20$)
and the velocity of the \ion{Si}{2}~$\lambda$6355 absorption
minimum is consistent with being constant in time from phases
as early as a week before, and up to two weeks after, $B$-band maximum light.
We interpret the velocity plateaus as evidence for a reverse-shock shell in
the ejecta formed by interaction at early times with a compact envelope of
surrounding material, as might be expected for SNe resulting from the mergers
of two white dwarfs.  We use the bolometric light curves and line velocity
evolution of these SNe to estimate important parameters
of the progenitor systems, including \nickel\ mass, total progenitor mass,
and masses of shells and surrounding carbon/oxygen envelopes.
We find that the reconstructed total progenitor mass distribution of the
events (including SN~2007if) is bounded from below by the Chandrasekhar mass,
with SN~2007if being the most massive.  We discuss the relationship of these
events to the emerging class of super-Chandrasekhar-mass SNe~Ia,
estimate the relative rates, compare the mass distribution to that
expected for double-degenerate SN~Ia progenitors from population synthesis,
and consider implications for future cosmological Hubble diagrams.
\end{abstract}


\vspace{0.1in}

\section{Introduction}

Type Ia supernovae (SNe~Ia) have become indispensable as luminosity distance
indicators for exploring the accelerated expansion of the universe
\citep{riess98,scp99}.  Their utility is due mainly to their very high
luminosities, in combination with a set of relations between intrinsic
luminosity, color and light curve width
\citep{riess98,tripp98,phillips99,goldhaber01}
which reduces their dispersion around the Hubble diagram to $\sim 0.15$~mag.
Much recent attention has been given to improving the precision of distance
measurements by searching for further standardization relations, with some
methods using near-maximum-light spectra to deliver core Hubble residual
dispersions as low as 0.12~mag \citep{sjb09,wang09,csp10,fk11}.

Despite ongoing research, however, many uncertainties remain regarding the
physical nature of SN~Ia progenitor systems,
although observational subclasses can be formed
\citep[e.g.,][]{bfn93,benetti05}.  Detailed observations of the light curve
of the nearby type Ia SN~2011fe shortly after explosion have shown that it
must have had a compact progenitor \citep[PTF11kly][]{nugent11,bloom12},
in line with expectations that SNe~Ia result from thermonuclear explosions
of white dwarfs in binary systems.  The mass and evolutionary state of the
progenitor's companion star, the events that trigger the explosion, and the
circumstellar and host galaxy environment of most normal SNe~Ia remain
unknown.  Next-generation SN~Ia cosmology experiments make stringent demands,
and any cosmological or astrophysical phenomenon which could
bias the measured luminosities of SNe~Ia at the level of a few percent has
become important to investigate \citep{kim04}.
If two or more progenitor channels exist corresponding to different peak
luminosities or luminosity standardization relations, evolution with redshift
of the relative rates of observed SNe~Ia from these channels could mimic
the effects of a time-varying dark energy equation of state \citep{linder06}.

The two main competing SN~Ia progenitor
scenarios are the \emph{single-degenerate} scenario \citep{wi73},
in which a carbon/oxygen white dwarf slowly accretes mass from a
non-degenerate companion until exploding near the Chandrasekhar mass,
and the \emph{double-degenerate} scenario \citep{it84}, in which two
white dwarfs collide or merge.  There are also \emph{sub-Chandrasekhar}
models \citep{ww94,vkcj10}, in which the explosion is triggered by the
detonation of a helium layer on the surface of a sub-Chandrasekhar-mass white
dwarf \citep{sim10}.  Although historically the single-degnerate scenario has
been favored, the double-degenerate scenario has recently gained ground
both theoretically and observationally.
\citet{gb10} used X-ray observations of nearby elliptical galaxies to set a
stringent limit on the number of accreting white dwarf systems in those
galaxies \citep[see also][]{rds10a}, and hence on the single-degenerate
contribution to the SN~Ia rate,
although their interpretation of the measurements has been questioned
\citep{rds10b,hkn10}.
Based on searches for an ex-companion star near the site of explosion,
\citet{li11b} ruled out a red giant companion for single-degenerate models
of SN~2011fe, and \citet{sp12} argue strongly that the supernova remnant
SNR~0509-67.5, in the Large Magellanic Cloud, must have had a
double-degenerate progenitor; however, single-degenerate scenarios have been
put forth in which long delays between formation of the primary white dwarf
and the SN~Ia explosion could allow the companion to evolve and become
fainter, evading attempts to detect them directly \citep{rds12}.
While a merging white dwarf system could also undergo accretion-induced
collapse to a neutron star \citep{nk91,sn98} rather than exploding as a
SN~Ia, theoretical investigations of SNe~Ia from mergers have also progressed.
Some merger simulations produce too much unburnt material to reproduce spectra
of normal SNe~Ia \citep{pfannes10a}; other simulations suggest that if the
merger process is violent enough to ignite the white dwarf promptly,
mergers may produce subluminous SNe~Ia \citep{pakmor11}
or even normal SNe~Ia \citep{pakmor12}.

Interest has also been aroused by the discovery of ``super-Chandra'' SNe~Ia
which far exceed the norm in luminosity:  SN~2003fg \citep{howell06},
SN~2006gz \citep{hicken07}, SN~2007if \citep{scalzo10,yuan10}, and SN~2009dc
\citep{yamanaka09,tanaka10,taub11,silverman11}.  If the light curves of
these SNe are powered by the radioactive decay of \nickel, as expected for
normal SNe~Ia, the \nickel\ mass necessary to produce the observed luminosity
implies a system mass significantly in excess of the Chandrasekhar mass.
The high luminosity of SN~2006gz reported in \citet{hicken07} hinges upon an
uncertain reddening correction, and late-phase photometry and
spectroscopy suggest a smaller \nickel\ mass than that inferred from the
dereddened peak luminosity \citep{maeda09a}; SN~2003fg, SN~2007if, and
SN~2009dc are much more luminous than normal SNe~Ia even before dereddening.
None of these SNe lie on the existing luminosity standardization relations.
If clear photometric and spectroscopic signatures of the progenitor system
or explosion mechanism can be discovered for these rare SNe~Ia, they may help
bring to light similar, but weaker, signatures in less-extreme SNe~Ia.

In addition to its high luminosity
($M_V = -20.4$), SN~2007if showed a red ($B-V = 0.18$) color at maximum light,
\ion{C}{2} absorption in spectra taken near maximum light, and a low
($\sim 8500$~km~s$^{-1}$) and very slowly-evolving \ion{Si}{2}~$\lambda 6355$
absorption velocity.  \citet{scalzo10} used this information to model
SN~2007if as a ``tamped detonation'' resulting from the explosion of a
super-Chandrasekhar-mass white dwarf inside a dense, but compact,
carbon/oxygen envelope,
as expected in some double-degenerate merger scenarios \citep{kmh93,hk96}.
Using the bolometric light curve between 60 and 120 days past explosion
to determine the optical depth for gamma-ray trapping in the ejecta
\citep[see e.g.][]{jeffery99,stritz06}, \citet{scalzo10} estimated the
total SN~2007if system mass to be $2.4 \pm 0.2~\Msol$, with $\sim 15\%$
of this mass bound up in the carbon/oxygen envelope formed in the merger
process.  Such a high mass is near the theoretical upper mass limit for
two carbon/oxygen white dwarfs in a binary system.
However, numerical calculations have since confirmed that the very large mass
of \nickel\ needed to explain the luminosity of SN~2007if can plausibly
be produced in a collision of two high-mass white dwarfs \citep{raskin10},
or in the prompt detonation of a single rapidly-rotating white dwarf
\citep{pfannes10b}.  Recent work has suggested that a white dwarf with a
non-degenerate companion could be spun up by accretion to high mass,
and remain rotationally supported for some time only to explode later
\citep{justham11,hachisu11}.

In contrast, the comparably-bright SN~2009dc had a normal color ($B-V = 0$)
near maximum light and showed rapid \ion{Si}{2} velocity evolution at
$\sim100$~km~s$^{-1}$~day$^{-1}$ \citep{silverman11}, difficult to explain
by a tamped detonation.  \citet{tanaka10} group SN~2009dc with
the 2003fg-like SNe~Ia based on its high luminosity and broad light curve,
but they compare it spectroscopically with SN~2006gz, given its strong
\ion{Si}{2}~$\lambda 6355$ and \ion{C}{2}~$\lambda 6580$ absorption at
early phases.  \citet{silverman11} also noted some spectroscopic differences
in the post-maximum spectra of SN~2007if and SN~2009dc.
Moreover, the SN was fainter at one year after explosion than expected for
the estimated \nickel\ mass \citep{taub11,silverman11}, as \citet{maeda09a}
noted for SN~2006gz.  \citet{taub11} calculated a total system mass of
$2.8~\Msol$ for SN~2009dc by estimating the diffusion time from the width
of the bolometric light curve \citep{arnett82};
they explored a number of different thermonuclear and
core-collapse explosion scenarios, and found none of them to be completely
satisfactory in explaining all the observations.

After the discovery of SN2007if, we remained vigilant for SNe with
similar characteristics; when possible, additional follow-up was obtained
when such SNe were recognized.  Given the high ionization state and
predominance of \ion{Fe}{2} and \ion{Fe}{3} absorption in SN~2007if's spectra
up until maximum light, characteristics shared with SN~1991T, we used
a 1991T-like spectroscopic classification to select for
super-Chandrasekhar-mass SN candidates, triggering follow-up
even for more distant, fainter examples.  The Nearby Supernova
Factory (SNfactory) obtained, as part of its spectroscopic follow-up program
on a large sample of nearby SNe~Ia, observations of five such
SNe~Ia besides SN~2007if, the presentation of which is the subject of
this paper.
Later examination of the full data set of SNfactory spectroscopic time series
showed that the SNe in our sample each also show a plateau in
the time evolution of the velocity of the \ion{Si}{2}~$\lambda 6355$
absorption minimum, lasting
from the earliest phase the velocity was measurable until 10--15 days after
maximum light, as in SN~2007if.  The absorption minimum velocities of other
intermediate-mass elements in these SNe also show plateau behavior.
These events have a different appearance from SN~2006gz and SN~2009dc,
which do not show velocity plateaus and show somewhat different behavior
in their early spectra and late-time light curves.  SN~2003fg was
spectroscopically observed only once, making it impossible to determine how
it evolved spectroscopically.

Our supernova discoveries, our sample selection, and the provenance of
our data are described in \S\ref{sec:observations}; the light curves and
spectra are presented in section \S\ref{sec:analysis}.
In \S\ref{sec:modeling} we model our SNe as tamped detonations, using the
formalism of \citet{scalzo10} to estimate an envelope mass and total system
mass for each SN.  We discuss the broader implications of our
results in \S\ref{sec:discussion}, including implications for progenitor
systems, explosion mechanisms, and cosmology,
and we summarize and conclude in \S\ref{sec:conclusions}.


\section{Observations}
\label{sec:observations}

\begin{deluxetable*}{lrrlcccc}
\tabletypesize{\footnotesize}
\tablecaption{Candidate super-Chandra SN discoveries from SNfactory\label{tbl:snsearch}}
\tablehead{
   \colhead{SN Name} &
   \colhead{RA} &
   \colhead{DEC} &
   \colhead{Disc. UT Date} &
   \colhead{Disc. Phase\tablenotemark{a}} &
   \colhead{$z_\mathrm{helio}$\tablenotemark{b}} &
   \colhead{Host Type\tablenotemark{c}} &
   \colhead{$E(B-V)_\mathrm{MW}$\tablenotemark{d}}
}
\startdata
SNF~20070528-003 & 16:47:31.46 & +21:28:33.4 & 2007 May 05 & $-7$
                 & 0.1171 & dIrr & 0.045 \\
SNF~20070803-005 & 22:26:24.03 & +21:14:56.6 & 2007 Aug 03.4 & $-11$
                 & 0.0315 & Sbc  & 0.047 \\
SN~2007if\tablenotemark{e} & 01:10:51.37 & +15:27:40.1 & 2007 Aug 25.4 & $-10$
                 & 0.0742 & dIrr & 0.082 \\
SNF~20070912-000 & 00:04:36.76 & +18:09:14.4 & 2007 Sep 12.4 & $-17$
                 & 0.1231 & Sbc  & 0.029 \\
SNF~20080522-000 & 13:36:47.59 & +05:08:30.4 & 2008 May 22 & $-18$
                 & 0.0453 & Sb   & 0.026 \\
SNF~20080723-012 & 16:16:03.26 & +03:03:17.4 & 2008 July 23.4 & $-18$
                 & 0.0745 & dIrr & 0.062 \\
\vspace{-0.1in}
\enddata
\tablenotetext{a}{In rest-frame days relative to $B$-band maximum light,
                  as determined from a SALT2 fit to the $K+S$-corrected
                  rest-frame light curve.}
\tablenotetext{b}{From template fit to host galaxy spectrum 
                  \citep[][M.~Childress \etal\ 2012, in preparation]
                  {childress11}.}
\tablenotetext{c}{Morpological type from visual inspection.}
\tablenotetext{d}{From \citet{sfd}.}
\tablenotetext{e}{Discovered independently by the Texas Supernova Search
                  \citep{cbet2007if,yuan10} and by SNfactory as
                  SNF~20070825-001 \citep{scalzo10}.}
\end{deluxetable*}

This section details the discovery, selection and follow-up data for our 
sample of candidate super-Chandrasekhar-mass SNe~Ia.

\subsection{Discovery}
\label{subsec:discovery}

The supernovae are among the 400 SNe~Ia discovered in the SNfactory
SN~Ia search, carried out between 2005 and 2008 with the QUEST-II camera
\citep{baltay07} mounted on the Samuel Oschin 1.2-m Schmidt telescope
at Palomar Observatory (``Palomar/QUEST'').  QUEST-II observations
were taken in a broad RG-610 filter with appreciable transmission
from 6100--10000~\AA, covering the Johnson $R$ and $I$ bandpasses.
Table~\ref{tbl:snsearch} lists the details of the SN discoveries,
including SN~2007if.

Upon discovery candidate SNe were spectroscopically screened using the
SuperNova Integral Field Spectrograph \citep[SNIFS;][]{snf,snifs} on the
University of Hawaii (UH) 2.2~m on Mauna Kea.  Our normal criteria for
continuing spectrophotometric follow-up of SNe~Ia with SNIFS were that
the spectroscopic phase be at or before maximum light, as estimated using
a template-matching code similar e.g. to SUPERFIT \citet{superfit},
and that the redshift be in the range $0.03<z<0.08$.
Aware of the potential for discovering nearby counterparts to the candidate
super-Chandrasekhar-mass SN~2003fg \cite{howell06}, we allowed exceptions
to our nominal redshift limit for continued follow-up if the spectrum
of a newly-screened candidate appeared unusual or especially early.

\subsection{Selection criteria}
\label{subsec:selection}

Our spectroscopic selection was informed by the existing spectra
of SN~2003fg and SN~2007if. In particular, the pre-maximum and
near-maximum spectra of SN~2007if showed weak \ion{Si}{2}~$\lambda 6355$
and \ion{Ca}{2}~H+K absorption and strong absorption from \ion{Fe}{2}
and \ion{Fe}{3}, with noted similarity to SN~1991T \citep{scalzo10}.
We therefore prioritized spectroscopic follow-up for SNe~Ia visually
similar to, or more extreme (i.e. having weaker IME and stronger Fe-peak
absorption) than, SN~1991T itself.  Here we have excluded the less
extreme 1999aa-likes, which can be separated based on the strength of
\ion{Ca}{2}~H+K \citep{silverman12}.  As a cross-check on our initial
selection conducted using the initial classification spectra, we have
run SNID~v5.0 \citep{snid} on the entire SNfactory sample, using version 1.0
of the templates supplemented by the \citet{scalzo10}
SNfactory spectra of SN~2007if.  We searched for
pre-maximum spectra for which the best subtype was ``Ia-91T'', or ``Ia-pec''
with the top match being a 1991T-like SN~Ia or SN~2007if itself;
this yielded the same set of SNe as the visual selection.

Our sample could thus be described as 1991T-like based on
their spectroscopic properties.  However, this categorization is often
used to imply lightcurve characteristics --- luminosity excess or slow
decline rate --- that were not part of our selection criteria.  Moreover,
since our interest is ultimately in the masses of these systems, we will
refer to these as ``candidate super-Chandra SNe~Ia''
throughout this paper.

The sample of six objects (including SN~2007if) presented here constitutes
all such spectroscopically-selected candidate super-Chandra SNe~Ia in
the SNfactory sample. This was established as part of the classification
cross-check described above.  An additional 141 SNe~Ia discovered by
the SNfactory were also followed spectrophotometrically and constitute
a homogenous comparison sample. Like SN~2003fg, but unlike SN~2006gz and
SN~2009dc, our sample of super-Chandra candidates and our reference sample
were discovered in a wide-area search and therefore sample the full range of
host galaxy environments. This may prove important in understanding the
formation of super-Chandra SNe~Ia, e.g., if metallicity plays an important
role \citep{taub11,khan11,hachisu11}.  We reiterate that characteristics such
as luminosity excess, velocity evolution, lightcurve shape, etc. were not
used in our selection --- it is purely based on optical spectrophotometry.


\subsection{Follow-up Observations}
\label{subsec:follow-up}

Most of the $BVRI$ photometry in this work was synthesized from
SNIFS flux-calibrated rest-frame spectra, using the bandpasses of
\citet{bessell90}, and corrected for Galactic dust extinction using
$E(B-V)$ from \citet{sfd} and the extinction law of \citet{cardelli} with
$R_V = 3.1$.  Follow-up $BVRI$ photometry for SNF 20070803-005, using
the ANDICAM imager on the CTIO 1.3-m, was obtained through the Small
and Moderate Aperture Research Telescope System (SMARTS) Consortium.
Redshifts were obtained from host galaxy spectra, which were either
extracted from the SNIFS datacubes, or taken separately using the Kast
Double Spectrograph \citep{kast} on the Shane 3~m telescope at Lick
Observatory, the Low Resolution Imaging Spectrograph \citep{lris}
at Keck-I on Mauna Kea or the Goodman High-Throughput Spectrograph 
at SOAR on Cerro Pachon \citep[see][and M.~Childress \etal\
2012, in preparation]{childress11}.  The redshifts (from spectroscopic
template fitting) and morphological types (from visual inspection)
are listed in Table~\ref{tbl:snsearch}.



\subsubsection{SNIFS Spectrophotometry}
\label{subsubsec:spectra_obs}

Observations of all six SNe were obtained with SNIFS,
built and operated by the SNfactory.
SNIFS is a fully integrated instrument optimized
for automated observation of point sources on a structured background
over the full optical window at moderate spectral resolution.  It
consists of a high-throughput wide-band pure-lenslet integral field
spectrograph \citep[IFS, ``\`a la
TIGER'';][]{bacon95,bacon00,bacon01}, a multifilter photometric
channel to image the field surrounding the IFS for atmospheric
transmission monitoring simultaneous with spectroscopy, and an
acquisition/guiding channel.  The IFS possesses a fully filled
$6\farcs 4 \times 6\farcs 4$ spectroscopic field of view (FOV) subdivided
into a grid of $15 \times 15$ spatial elements (spaxels), a
dual-channel spectrograph covering 3200--5200~\AA\ and
5100--10000~\AA\ simultaneously, and an internal calibration unit
(continuum and arc lamps).  SNIFS is continuously mounted on the south
bent Cassegrain port of the UH 2.2~m telescope
(Mauna Kea) and is operated remotely.
The SNIFS spectrophotometric data reduction pipeline has been
described in previous papers
\citep{bacon01,snf2005gj,scalzo10}.
We subtract the host galaxy light using the methodology
described in \citet{bongard11}, which uses SNIFS IFU exposures of the host
taken after each SN has faded away.


\subsubsection{SMARTS Photometry}
\label{subsubsec:bvri_obs}

The SMARTS imaging of SNF~20070803-005 and SN~2007if with ANDICAM
consisted of multiple 240~s exposures in each of the $BVRI$ filters at each
epoch.  The images were processed using an automated pipeline based on
IRAF \citep{iraf}; this pipeline was used in \citet{scalzo10} and is
described in detail there.  We briefly summarize the methods below.

All ANDICAM images were bias-subtracted, overscan-subtracted, and flat-fielded
by the SMARTS Consortium, also using IRAF (\texttt{ccdproc}).
In each band, four to six final reference images of each SN field were taken
at least a year after explosion,
and combined to form a co-add.  This co-add was then registered, normalized,
and convolved to match the observing conditions of each SN image, before
subtraction to remove the host galaxy light.


An absolute calibration (zeropoint, extinction and color terms)
was established on photometric nights from observations of \cite{landolt92}
standards, fitting a zeropoint and extinction coefficient for each night
separately as well as a color term constant across all nights.
The calibration was transferred to the field stars for each photometric
night separately using the zeropoint and extinction but ignoring the color
terms, producing magnitudes on a ``natural'' ANDICAM system which agrees
with the Landolt system for stars with $B-V = V-R = R-I = 0$.
These calibrated magnitudes were then averaged over photometric nights
to produce final calibrated ANDICAM magnitudes for the field stars.

Each SN's ANDICAM-system $BVRI$ light curve was then measured by comparison
to the field stars.  To fix the SN's location, we found the mean position
over observations in the same filter, weighting by the signal-to-noise ratio
(S/N) of each detection.  We then measured the final flux in each image in
a circular aperture centered at this mean location.  The observer-frame
ANDICAM magnitudes were corrected for Galactic extinction using $E(B-V)$
from \citet{sfd} and the extinction law of \citet{cardelli} with $R_V = 3.1$.
The SN magnitudes were $K$-corrected \citep{nkp02} to rest-frame Bessell
$BVRI$ bandpasses \citep{bessell90} using the SNIFS spectrophotometric time
series and a set of ANDICAM system throughput curves, the central wavelengths
of which were shifted to match ANDICAM observations of spectrophotometric
standard stars to published synthetic photometry \citep{stritz05}.

\subsection{Lightcurves}
\label{subsec:lightcurves}

\begin{figure*}
\center
\resizebox{\textwidth}{!}{\includegraphics{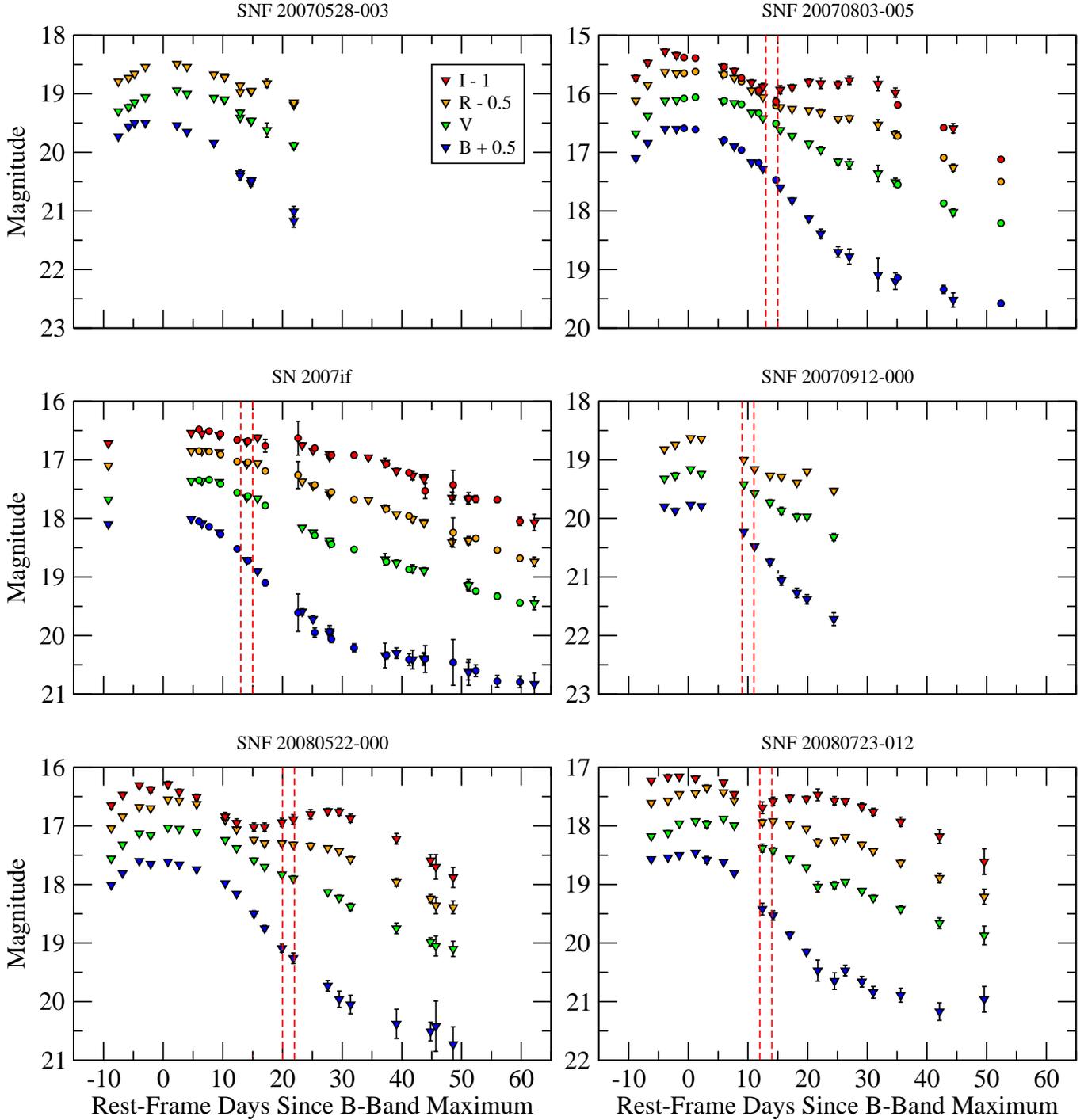}}
\caption{\small Rest-frame Bessell $BVRI$ light curves from ANDICAM+SNIFS.
Inverted triangles:  Rest-frame Bessell $BVRI$ magnitudes synthesized
from SNIFS flux-calibrated spectroscopy.
Circles:  ANDICAM $BVRI$, $K$-corrected to the respective rest-frame
Bessell filters.
Vertical dotted red lines mark the light curve phase range corresponding to
the break in \vSi discussed in \S\ref{subsec:specanal}.}
\label{fig:bvrilc}
\end{figure*}

\begin{figure*}
\center
\resizebox{\textwidth}{!}{\includegraphics{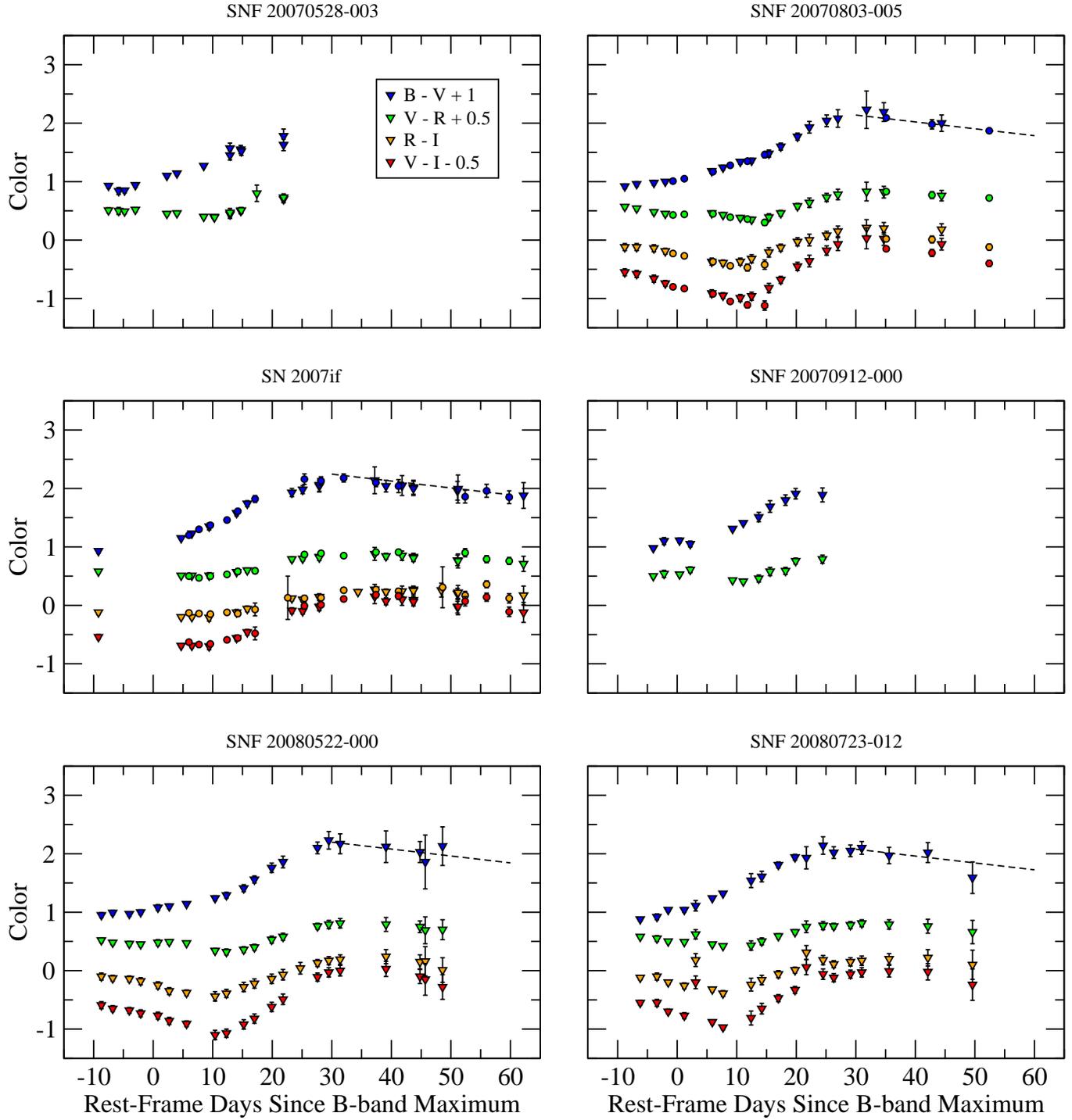}}
\caption{\small Rest-frame Bessell color evolution
from ANDICAM (circles) and SNIFS spectrophotometry (inverted triangles).
Dotted lines indicate fits to the Lira relation with a floating excess.}
\label{fig:bvlira}
\end{figure*}

The rest-frame, Milky Way de-reddened Bessell $BVRI$ light curves of the
SNfactory super-Chandra candidates are given in Table~\ref{tbl:bvrilc}
and shown in Figure~\ref{fig:bvrilc}.
The color evolution is shown in Figure~\ref{fig:bvlira}.
The light curves of the two SNe at the high end of the SNfactory redshift
range, SNF~20070528-003 ($z = 0.117$) and SNF~20070912-000 ($z = 0.123$),
have less extensive coverage than the other SNe; the S/N is
lower, and a significant fraction of the rest-frame $I$-band
transmission lies outside of the observer-frame
wavelength range of the SNIFS spectrograph.
For a small number of observations, rest-frame $B$-band or $V$-band
measurements are unavailable due to instrument problems 
with the SNIFS blue channel.  The detailed analysis
of these lightcurves is presented in \S\ref{subsec:reddening} and
\S\ref{subsec:bololc}.


\begin{deluxetable*}{lrccccc}
\tabletypesize{\footnotesize}
\tablecaption{Rest-frame $BVRI$ light curves\label{tbl:bvrilc}}
\tablehead{
   \colhead{MJD\tablenotemark{a}} &
   \colhead{Phase\tablenotemark{b}} &
   \colhead{$B$} &
   \colhead{$V$} &
   \colhead{$R$} &
   \colhead{$I$} &
   \colhead{Instrument}
}
\startdata
% ...
\cutinhead{SNF~20070528-003}
54250.6 & $-7.5$ & $19.23 \pm 0.03$ & $19.30 \pm 0.03$ & $19.29 \pm 0.02$ &          \nodata & SNIFS \\*
54252.5 & $-5.8$ & $19.06 \pm 0.04$ & $19.23 \pm 0.05$ & $19.23 \pm 0.03$ &          \nodata & SNIFS \\*
54253.5 & $-4.8$ & $19.00 \pm 0.03$ & $19.15 \pm 0.04$ & $19.16 \pm 0.03$ &          \nodata & SNIFS \\*
54255.5 & $-3.0$ & $19.00 \pm 0.02$ & $19.06 \pm 0.03$ & $19.04 \pm 0.02$ &          \nodata & SNIFS \\*
54261.5 & $ 2.3$ & $19.04 \pm 0.02$ & $18.94 \pm 0.02$ & $18.99 \pm 0.02$ &          \nodata & SNIFS \\*
54263.4 & $ 4.0$ & $19.15 \pm 0.02$ & $19.00 \pm 0.02$ & $19.04 \pm 0.02$ &          \nodata & SNIFS \\*
54268.4 & $ 8.5$ & $19.34 \pm 0.02$ & $19.07 \pm 0.02$ & $19.17 \pm 0.02$ &          \nodata & SNIFS \\*
54270.4 & $10.2$ &          \nodata & $19.10 \pm 0.02$ & $19.20 \pm 0.02$ &          \nodata & SNIFS \\*
54270.4 & $10.3$ &          \nodata & $19.10 \pm 0.02$ & $19.22 \pm 0.02$ &          \nodata & SNIFS \\*
54273.3 & $12.9$ & $19.86 \pm 0.07$ & $19.41 \pm 0.05$ & $19.47 \pm 0.04$ &          \nodata & SNIFS \\*
54273.4 & $12.9$ & $19.90 \pm 0.07$ & $19.32 \pm 0.05$ & $19.36 \pm 0.04$ &          \nodata & SNIFS \\*
54275.4 & $14.7$ & $20.02 \pm 0.06$ & $19.47 \pm 0.04$ & $19.46 \pm 0.03$ &          \nodata & SNIFS \\*
54275.5 & $14.8$ & $19.98 \pm 0.05$ & $19.46 \pm 0.04$ & $19.45 \pm 0.03$ &          \nodata & SNIFS \\*
54278.4 & $17.4$ &          \nodata & $19.62 \pm 0.12$ & $19.32 \pm 0.07$ &          \nodata & SNIFS \\*
54283.4 & $21.9$ & $20.67 \pm 0.11$ & $19.89 \pm 0.05$ & $19.69 \pm 0.03$ &          \nodata & SNIFS \\*
54283.4 & $21.9$ & $20.51 \pm 0.09$ & $19.88 \pm 0.05$ & $19.65 \pm 0.03$ &          \nodata & SNIFS \\[-0.05in]

% ...
\cutinhead{SNF~20070803-005}
54318.5 & $-8.8$ & $16.60 \pm 0.03$ & $16.68 \pm 0.03$ & $16.62 \pm 0.03$ & $16.73 \pm 0.05$ & SNIFS \\*
54320.5 & $-6.8$ & $16.34 \pm 0.03$ & $16.38 \pm 0.03$ & $16.35 \pm 0.03$ & $16.47 \pm 0.05$ & SNIFS \\*
54323.5 & $-3.9$ & $16.10 \pm 0.02$ & $16.12 \pm 0.03$ & $16.13 \pm 0.03$ & $16.28 \pm 0.05$ & SNIFS \\*
54325.5 & $-2.0$ & $16.10 \pm 0.02$ & $16.11 \pm 0.03$ & $16.15 \pm 0.03$ & $16.34 \pm 0.05$ & SNIFS \\*
54326.8 & $-0.7$ & $16.09 \pm 0.01$ & $16.08 \pm 0.01$ & $16.15 \pm 0.02$ & $16.38 \pm 0.02$ & SMARTS  \\*
54328.7 & $ 1.2$ & $16.11 \pm 0.01$ & $16.06 \pm 0.01$ & $16.12 \pm 0.02$ & $16.39 \pm 0.02$ & SMARTS  \\*
54333.5 & $ 5.8$ & $16.31 \pm 0.03$ & $16.13 \pm 0.03$ & $16.17 \pm 0.03$ & $16.54 \pm 0.06$ & SNIFS \\*
54333.7 & $ 6.0$ & $16.29 \pm 0.01$ & $16.12 \pm 0.02$ & $16.17 \pm 0.03$ & $16.54 \pm 0.02$ & SMARTS  \\*
54335.5 & $ 7.7$ & $16.40 \pm 0.03$ & $16.16 \pm 0.02$ & $16.23 \pm 0.02$ & $16.61 \pm 0.05$ & SNIFS \\*
54336.7 & $ 8.9$ & $16.46 \pm 0.02$ & $16.18 \pm 0.02$ & $16.29 \pm 0.02$ & $16.73 \pm 0.02$ & SMARTS  \\*
54338.5 & $10.6$ & $16.67 \pm 0.03$ & $16.32 \pm 0.03$ & $16.44 \pm 0.03$ & $16.81 \pm 0.05$ & SNIFS \\*
54339.7 & $11.8$ & $16.68 \pm 0.02$ & $16.33 \pm 0.02$ & $16.47 \pm 0.03$ & $16.94 \pm 0.05$ & SMARTS  \\*
54340.4 & $12.5$ & $16.78 \pm 0.03$ & $16.42 \pm 0.03$ & $16.56 \pm 0.03$ & $16.88 \pm 0.06$ & SNIFS \\*
54342.7 & $14.7$ & $16.97 \pm 0.03$ & $16.51 \pm 0.04$ & $16.70 \pm 0.04$ & $17.13 \pm 0.07$ & SMARTS  \\*
54343.5 & $15.4$ & $17.10 \pm 0.04$ & $16.62 \pm 0.04$ & $16.73 \pm 0.04$ & $16.93 \pm 0.07$ & SNIFS \\*
54345.5 & $17.4$ & $17.32 \pm 0.04$ & $16.72 \pm 0.03$ & $16.76 \pm 0.03$ & $16.89 \pm 0.05$ & SNIFS \\*
54348.4 & $20.2$ & $17.63 \pm 0.05$ & $16.85 \pm 0.04$ & $16.78 \pm 0.04$ & $16.80 \pm 0.06$ & SNIFS \\*
54350.4 & $22.2$ & $17.89 \pm 0.08$ & $16.96 \pm 0.06$ & $16.82 \pm 0.06$ & $16.82 \pm 0.09$ & SNIFS \\*
54353.4 & $25.1$ & $18.20 \pm 0.09$ & $17.16 \pm 0.05$ & $16.93 \pm 0.04$ & $16.84 \pm 0.06$ & SNIFS \\*
54355.4 & $27.0$ & $18.28 \pm 0.13$ & $17.20 \pm 0.08$ & $16.92 \pm 0.05$ & $16.77 \pm 0.07$ & SNIFS \\*
54360.3 & $31.8$ & $18.59 \pm 0.28$ & $17.36 \pm 0.14$ & $17.03 \pm 0.09$ & $16.83 \pm 0.12$ & SNIFS \\*
54363.3 & $34.7$ & $18.70 \pm 0.14$ & $17.51 \pm 0.07$ & $17.19 \pm 0.06$ & $16.98 \pm 0.08$ & SNIFS \\*
54363.7 & $35.1$ & $18.64 \pm 0.03$ & $17.55 \pm 0.02$ & $17.22 \pm 0.02$ & $17.19 \pm 0.02$ & SMARTS  \\*
54371.7 & $42.8$ & $18.84 \pm 0.07$ & $17.87 \pm 0.04$ & $17.59 \pm 0.04$ & $17.58 \pm 0.04$ & SMARTS  \\*
54373.3 & $44.4$ & $19.02 \pm 0.12$ & $18.02 \pm 0.06$ & $17.76 \pm 0.06$ & $17.59 \pm 0.08$ & SNIFS \\*
54381.6 & $52.4$ & $19.08 \pm 0.03$ & $18.21 \pm 0.02$ & $18.00 \pm 0.02$ & $18.12 \pm 0.04$ & SMARTS  \\[-0.05in]

% ...
\cutinhead{SNF~20070912-000}
54358.4 & $-4.0$ & $19.30 \pm 0.02$ & $19.32 \pm 0.02$ & $19.32 \pm 0.02$ &          \nodata & SNIFS \\*
54360.4 & $-2.2$ & $19.37 \pm 0.04$ & $19.27 \pm 0.05$ & $19.24 \pm 0.03$ &          \nodata & SNIFS \\*
54363.4 & $ 0.4$ & $19.27 \pm 0.02$ & $19.16 \pm 0.03$ & $19.13 \pm 0.02$ &          \nodata & SNIFS \\*
54365.4 & $ 2.2$ & $19.29 \pm 0.03$ & $19.24 \pm 0.04$ & $19.14 \pm 0.03$ &          \nodata & SNIFS \\*
54373.4 & $ 9.3$ & $19.73 \pm 0.02$ & $19.42 \pm 0.02$ & $19.50 \pm 0.02$ &          \nodata & SNIFS \\*
54375.4 & $11.1$ & $19.98 \pm 0.03$ & $19.57 \pm 0.02$ & $19.66 \pm 0.02$ &          \nodata & SNIFS \\*
54378.3 & $13.7$ & $20.24 \pm 0.06$ & $19.73 \pm 0.05$ & $19.77 \pm 0.04$ &          \nodata & SNIFS \\*
54380.5 & $15.6$ & $20.56 \pm 0.08$ & $19.87 \pm 0.06$ & $19.79 \pm 0.04$ &          \nodata & SNIFS \\*
54383.4 & $18.2$ & $20.77 \pm 0.08$ & $19.97 \pm 0.05$ & $19.89 \pm 0.03$ &          \nodata & SNIFS \\*
54385.3 & $19.9$ & $20.88 \pm 0.08$ & $19.97 \pm 0.04$ & $19.70 \pm 0.03$ &          \nodata & SNIFS \\*
54390.3 & $24.4$ & $21.22 \pm 0.11$ & $20.32 \pm 0.06$ & $20.03 \pm 0.04$ &          \nodata & SNIFS \\*
54390.4 & $24.4$ &          \nodata & $20.34 \pm 0.06$ & $20.06 \pm 0.03$ &          \nodata & SNIFS \\[-0.05in]

% ...
\vspace{-0.05in}
\enddata
\tablenotetext{a}{Observer frame $\mathrm{JD} - 2400000.5$.}
\tablenotetext{b}{In rest-frame days relative to $B$-band maximum light.}
\end{deluxetable*}

\begin{deluxetable*}{lrccccc}
\addtocounter{table}{-1}
\tabletypesize{\footnotesize}
\tablecaption{Rest-frame $BVRI$ light curves, continued}
\tablehead{
   \colhead{MJD\tablenotemark{a}} &
   \colhead{Phase\tablenotemark{b}} &
   \colhead{$B$} &
   \colhead{$V$} &
   \colhead{$R$} &
   \colhead{$I$} &
   \colhead{Instrument}
}
\startdata
% ...
\cutinhead{SNF~20080522-000}
54612.4 & $-8.7$ & $17.51 \pm 0.03$ & $17.56 \pm 0.03$ & $17.54 \pm 0.03$ & $17.65 \pm 0.05$ & SNIFS \\*
54614.4 & $-6.8$ & $17.31 \pm 0.02$ & $17.32 \pm 0.03$ & $17.34 \pm 0.03$ & $17.47 \pm 0.04$ & SNIFS \\*
54617.3 & $-4.0$ & $17.10 \pm 0.02$ & $17.13 \pm 0.03$ & $17.18 \pm 0.03$ & $17.31 \pm 0.04$ & SNIFS \\*
54619.3 & $-2.1$ & $17.15 \pm 0.03$ & $17.16 \pm 0.03$ & $17.20 \pm 0.03$ & $17.38 \pm 0.05$ & SNIFS \\*
54622.3 & $ 0.8$ & $17.11 \pm 0.03$ & $17.03 \pm 0.04$ & $17.05 \pm 0.03$ & $17.29 \pm 0.05$ & SNIFS \\*
54624.3 & $ 2.7$ & $17.16 \pm 0.03$ & $17.05 \pm 0.03$ & $17.07 \pm 0.03$ & $17.42 \pm 0.05$ & SNIFS \\*
54627.3 & $ 5.6$ & $17.24 \pm 0.02$ & $17.10 \pm 0.02$ & $17.13 \pm 0.02$ & $17.51 \pm 0.05$ & SNIFS \\*
54632.3 & $10.4$ & $17.48 \pm 0.03$ & $17.24 \pm 0.03$ & $17.40 \pm 0.03$ & $17.84 \pm 0.07$ & SNIFS \\*
54634.3 & $12.3$ & $17.66 \pm 0.03$ & $17.38 \pm 0.03$ & $17.56 \pm 0.03$ & $17.95 \pm 0.07$ & SNIFS \\*
54637.3 & $15.2$ & $18.00 \pm 0.05$ & $17.59 \pm 0.04$ & $17.74 \pm 0.04$ & $18.02 \pm 0.07$ & SNIFS \\*
54639.3 & $17.0$ & $18.25 \pm 0.05$ & $17.70 \pm 0.03$ & $17.80 \pm 0.04$ & $18.02 \pm 0.07$ & SNIFS \\*
54642.3 & $19.9$ & $18.59 \pm 0.07$ & $17.83 \pm 0.04$ & $17.80 \pm 0.04$ & $17.94 \pm 0.07$ & SNIFS \\*
54644.3 & $21.8$ & $18.76 \pm 0.09$ & $17.90 \pm 0.05$ & $17.82 \pm 0.04$ & $17.89 \pm 0.08$ & SNIFS \\*
54647.3 & $24.7$ &          \nodata &          \nodata & $17.84 \pm 0.05$ & $17.80 \pm 0.08$ & SNIFS \\*
54650.3 & $27.6$ & $19.23 \pm 0.09$ & $18.13 \pm 0.04$ & $17.88 \pm 0.03$ & $17.75 \pm 0.05$ & SNIFS \\*
54652.3 & $29.5$ & $19.46 \pm 0.14$ & $18.23 \pm 0.05$ & $17.93 \pm 0.04$ & $17.76 \pm 0.06$ & SNIFS \\*
54654.3 & $31.4$ & $19.55 \pm 0.16$ & $18.38 \pm 0.06$ & $18.07 \pm 0.05$ & $17.87 \pm 0.07$ & SNIFS \\*
54662.3 & $39.1$ & $19.88 \pm 0.25$ & $18.75 \pm 0.09$ & $18.46 \pm 0.07$ & $18.22 \pm 0.09$ & SNIFS \\*
54668.3 & $44.8$ & $20.01 \pm 0.16$ & $18.98 \pm 0.07$ & $18.74 \pm 0.07$ & $18.59 \pm 0.10$ & SNIFS \\*
54669.3 & $45.7$ & $19.92 \pm 0.43$ & $19.05 \pm 0.17$ & $18.86 \pm 0.14$ & $18.70 \pm 0.21$ & SNIFS \\*
54672.3 & $48.6$ & $20.23 \pm 0.30$ & $19.10 \pm 0.13$ & $18.89 \pm 0.11$ & $18.88 \pm 0.17$ & SNIFS \\[-0.05in]

% ...
\cutinhead{SNF~20080723-012}
54674.3 & $-6.2$ & $18.07 \pm 0.02$ & $18.18 \pm 0.02$ & $18.11 \pm 0.02$ & $18.23 \pm 0.03$ & SNIFS \\*
54677.3 & $-3.4$ & $18.04 \pm 0.03$ & $18.12 \pm 0.04$ & $18.07 \pm 0.03$ & $18.17 \pm 0.05$ & SNIFS \\*
54679.3 & $-1.5$ & $18.00 \pm 0.02$ & $17.96 \pm 0.02$ & $17.96 \pm 0.02$ & $18.16 \pm 0.03$ & SNIFS \\*
54682.3 & $ 1.2$ & $17.96 \pm 0.02$ & $17.92 \pm 0.03$ & $17.94 \pm 0.02$ & $18.19 \pm 0.04$ & SNIFS \\*
54684.4 & $ 3.1$ & $18.08 \pm 0.06$ & $17.97 \pm 0.06$ & $17.85 \pm 0.05$ & $17.67 \pm 0.09$ & SNIFS \\*
54687.3 & $ 5.9$ & $18.12 \pm 0.02$ & $17.88 \pm 0.02$ & $17.93 \pm 0.02$ & $18.26 \pm 0.04$ & SNIFS \\*
54689.3 & $ 7.7$ & $18.31 \pm 0.02$ & $17.99 \pm 0.02$ & $18.07 \pm 0.02$ & $18.46 \pm 0.04$ & SNIFS \\*
54694.3 & $12.4$ & $18.92 \pm 0.10$ & $18.38 \pm 0.07$ & $18.44 \pm 0.05$ & $18.69 \pm 0.10$ & SNIFS \\*
54696.3 & $14.2$ & $19.03 \pm 0.08$ & $18.42 \pm 0.05$ & $18.42 \pm 0.03$ & $18.58 \pm 0.07$ & SNIFS \\*
54699.3 & $17.0$ & $19.36 \pm 0.05$ & $18.56 \pm 0.03$ & $18.47 \pm 0.03$ & $18.52 \pm 0.05$ & SNIFS \\*
54702.3 & $19.8$ & $19.65 \pm 0.04$ & $18.71 \pm 0.03$ & $18.55 \pm 0.03$ & $18.54 \pm 0.05$ & SNIFS \\*
54704.3 & $21.7$ & $19.97 \pm 0.18$ & $19.04 \pm 0.09$ & $18.78 \pm 0.06$ & $18.47 \pm 0.10$ & SNIFS \\*
54707.3 & $24.5$ & $20.15 \pm 0.14$ & $19.01 \pm 0.06$ & $18.75 \pm 0.04$ & $18.57 \pm 0.07$ & SNIFS \\*
54709.3 & $26.3$ & $19.97 \pm 0.09$ & $18.96 \pm 0.04$ & $18.69 \pm 0.03$ & $18.58 \pm 0.05$ & SNIFS \\*
54712.3 & $29.1$ & $20.16 \pm 0.09$ & $19.11 \pm 0.04$ & $18.82 \pm 0.04$ & $18.67 \pm 0.06$ & SNIFS \\*
54714.3 & $31.0$ & $20.34 \pm 0.10$ & $19.23 \pm 0.05$ & $18.93 \pm 0.04$ & $18.76 \pm 0.06$ & SNIFS \\*
54719.3 & $35.6$ & $20.39 \pm 0.12$ & $19.42 \pm 0.06$ & $19.13 \pm 0.05$ & $18.93 \pm 0.08$ & SNIFS \\*
54726.3 & $42.1$ & $20.67 \pm 0.15$ & $19.66 \pm 0.09$ & $19.39 \pm 0.08$ & $19.18 \pm 0.12$ & SNIFS \\*
54734.2 & $49.6$ & $20.46 \pm 0.22$ & $19.87 \pm 0.16$ & $19.71 \pm 0.13$ & $19.61 \pm 0.22$ & SNIFS \\[-0.05in]

% ...
\vspace{-0.05in}
\enddata
\tablenotetext{a}{Observer frame $\mathrm{JD} - 2400000.5$.}
\tablenotetext{b}{In rest-frame days relative to $B$-band maximum light.}
\end{deluxetable*}


\subsection{Spectra}
\label{subsec:spectra}

\begin{figure*}
\center
\resizebox{5.5in}{!}
{\includegraphics{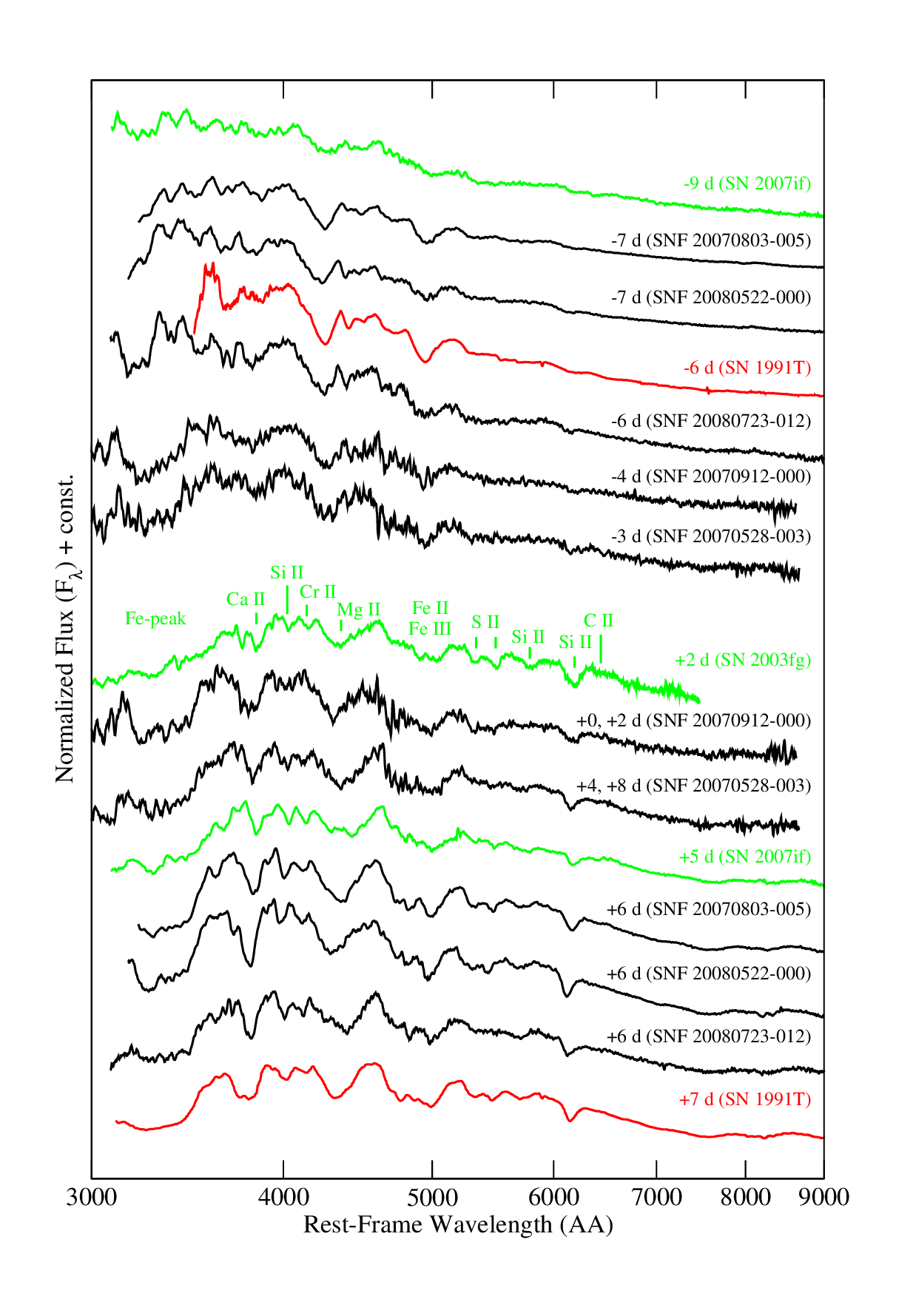}}
\caption{\small Selected spectra near maximum light of SNfactory candidate
super-Chandra SNe~Ia (black), with spectra of SN~1991T (red)
at $-6$~d \citep{mdt95} and at $+7$~d \citep{fil92},
and the candidate super-Chandra events SN~2003fg \citep[][green]{howell06}
and SN~2007if \citep[][green]{scalzo10} for comparison.
Spectra have been smoothed with a 15-point (width 2000 km~s$^{-1}$)
Savitzsky-Golay filter for presentation purposes.
Consecutive post-maximum spectra of the $z \sim 0.12$ SNe SNF~20070528-003
and SNF~20070912-000 have been co-added to improve signal-to-noise.
Common features identified with
\texttt{SYNAPPS} \citep{synapps} are marked.}
\label{fig:spectra-max}
\end{figure*}

Figure~\ref{fig:spectra-max}
presents the subset of spectra that were taken near maximum light for the six SNfactory
SNe~Ia, including SN~2007if.
Before maximum light, the spectra show weak \ion{Si}{2}, \ion{S}{2},
and \ion{Ca}{2} absorption plus strong \ion{Fe}{2} and \ion{Fe}{3} absorption,
in accordance with our selection criteria.
SN~1991T and the prototype candidate super-Chandra SN~Ia~2003fg
are shown for comparison.
After maximum light, some features noted in SN~2003fg and SN~2007if appear,
including the iron-peak blend near 4500~\AA, the sharp notch near
4130~\AA\ identified as \ion{Cr}{2} in \citet{scalzo10}
\citep[or tentatively as \ion{C}{2}~$\lambda 4237$ in][]{howell06}
and the blended lines near 3300~\AA\ identified as \ion{Cr}{2} and \ion{Co}{2}
in \citet{scalzo10}.  (The 4130~\AA\ feature is not clearly present in
the $z = 0.123$ candidate SNF~20070912-000, although this may be due to
the lower S/N of the spectrum.)  \citet{taub11} note that a similar feature in
SN~2009dc strengthens with time rather than fading, as one would expect for
\ion{Cr}{2} rather than \ion{C}{2}.

SN~2007if also shows a weak \ion{C}{2}~$\lambda 6580$ line in the post-maximum
spectra, which \citet{scalzo10} interpreted as a signature of unburned
material from the explosion.  While this line is not detected unambiguously
in the other SNe, the unusually shallow slope of the red wing of
\ion{Si}{2}~$\lambda 6355$ e.g. in SNF~20070528-003 and SNF~20080723-012 may
be a sign that \ion{C}{2} is present in these SNe \citep{rct11}.
The spectral properties of the full dataset are analyzed in \S\ref{subsec:specanal}.


\section{Analysis}
\label{sec:analysis}

The analysis (this section), modeling (\S\ref{sec:modeling}) and
interpretation (\S\ref{sec:discussion}) for our sample parallels that
made for SN~2007if in \citet{scalzo10}, with improvements described below.
We include a re-analysis of our observations of SN~2007if in this paper using
the improved techniques, for a more direct comparison with the
other SNe presented in this paper.


\subsection{Maximum-Light Behavior, Colors and Extinction}
\label{subsec:reddening}

\begin{deluxetable*}{lrrrrrrrrr}
\tabletypesize{\footnotesize}
\tablecaption{Derived quantities from SALT2 fits to light curves
              \label{tbl:salt2fit}}
\tablehead{
   \colhead{SN Name} &
   \colhead{MJD($B_\mathrm{max}$)} &
   \colhead{$M_V$\tablenotemark{a}} &
   \colhead{$x_1$} &
   \colhead{$c$} &
   \colhead{$\chi^2/\nu$} &
   \colhead{$\Delta m_{15}(B)$\tablenotemark{b}} &
   \colhead{$(B-V)_\mathrm{max}$} &
   \colhead{$B_\mathrm{max}-V_\mathrm{max}$} &
   \colhead{$\Delta mu$\tablenotemark{c}}
}
\startdata
SNF~20070528-003 & $54258.7$ & $-19.71 \pm 0.02$
                 & $1.08 \pm 0.20$ & $0.02 \pm 0.03$ & 1.82 & $0.84 \pm 0.05$
                 & $0.01 \pm 0.04$ & $0.03 \pm 0.04$ & $-0.48$ \\
SNF~20070803-005 & $54327.6$ & $-19.59 \pm 0.04$
                 & $1.04 \pm 0.09$ & $0.01 \pm 0.03$ & 0.74 & $0.85 \pm 0.05$
                 & $0.00 \pm 0.04$ & $0.02 \pm 0.04$ & $-0.28$ \\
SN~2007if        & $54346.8$ & $-20.38 \pm 0.02$
                 & $1.41 \pm 0.13$ & $0.12 \pm 0.03$ & 3.48 & $0.79 \pm 0.06$
                 & $0.12 \pm 0.05$ & $0.14 \pm 0.04$ & $-1.34$ \\
SNF~20070912-000 & $54363.0$ & $-19.69 \pm 0.02$
                 & $-0.03 \pm 0.19$ & $0.04 \pm 0.03$ & 1.41 & $1.08 \pm 0.04$
                 & $0.05 \pm 0.02$ & $0.06 \pm 0.02$ & $-0.52$ \\
SNF~20080522-000 & $54621.4$ & $-19.59 \pm 0.03$
                 & $1.18 \pm 0.13$ & $-0.01 \pm 0.03$ & 0.43 & $0.83 \pm 0.06$
                 & $-0.02 \pm 0.04$ & $0.00 \pm 0.04$ & $-0.17$ \\
SNF~20080723-012 & $54680.9$ & $-19.71 \pm 0.02$
                 & $0.53 \pm 0.12$ & $0.04 \pm 0.03$ & 2.61 & $0.93 \pm 0.04$
                 & $0.03 \pm 0.03$ & $0.05 \pm 0.03$ & $-0.59$ \\
\vspace{-0.1in}
\enddata
\tablenotetext{a}{Evaluated from SALT2 rest-frame $V_\mathrm{max}$,
                  with distance modulus at the host galaxy redshift assuming
                  a \lcdm\ cosmology with $\Omega_\mathrm{M} = 0.28$,
                  $\Omega_\Lambda = 0.72$, $H_0 = 72$~\kms~Mpc$^{-1}$.}
\tablenotetext{b}{Light curve decline rate, evaluated directly from
                  the best-fit SALT2 model, accounting for error in
                  the date of $B$-band maximum light.}
\tablenotetext{c}{Hubble residuals from \lcdm\ cosmology, evaluated
                  using equation~2 of \cite{snls3yr-cosmo}.}
\end{deluxetable*}

\newcommand{\na}{\ion{Na}{1}~D}
\newcommand{\ewna}{\ensuremath{EW}(\ion{Na}{1}~D)}

As discussed in \citet{scalzo10}, SN~2007if shows no distinct second maximum.
While Figure \ref{fig:bvrilc} does show second maxima for our other SNe,
its prominence is suppressed relative to normal SNe~Ia.  In SNF~20070803-005
and SNF~20080723-012, the peak-to-trough difference in a quintic polynomial
fit to the data from day +7 to day +42 is only 0.11~mag,
vs. 0.23~mag for SNF~20080522-000 and 0.35~mag for the SALT2 model with
$x1 = 1$, $c = 0$.  \citet{kasen06} noted three physical effects which could
reduce the contrast of the $I$-band second maximum:  low \nickel\ mass,
efficient mixing of \nickel\ into the outer layers of ejecta, and greater
absorption in the \ion{Ca}{2}~NIR triplet line source function.
Since our SNe are all overluminous with broad light curves, Arnett's rule
gives a high nickel mass, as we find in the next section.  The $I$-band
\emph{first} maximum is roughly concurrent with $B$-band maximum for our
SNe, rather than being significantly delayed
\citep[see figure 14 of][]{kasen06}, so it seems unlikely that emission in
the \ion{Ca}{2}~NIR triplet is contributing significantly.  The most likely
interpretation, especially given the prominence of Fe-peak lines in early
spectra of our SNe, is that \nickel\ is well-mixed into the outer layers.

Following practice from \cite{scalzo10}, we use the updated version (v2.2)
of the SALT2 light curve fitter \citep{guy10} to interpolate the magnitudes
and colors of each SN around maximum light, and to establish a date of
$B$-band maximum with respect to which we can measure light-curve phase.
While we use SALT2 here as a convenient functional form for describing
the shape of the light curves near maximum light, and to extract the usual
parameters describing the light curve shape, we do not expect the SALT2
model, trained on normal SNe~Ia, to give robust predictions for these
peculiar SNe~Ia outside the phase and wavelength coverage for each SN.
To minimize the impact of details of the SALT2 spectral model on the outcome,
we use SALT2 in the rest frame, include SALT2 light curve model errors in
the fitting, and we fit $BVR$ bands only; $I$-band is excluded from the fit.
The quantities derived from the SALT2 light curve fits are shown in
Table~\ref{tbl:salt2fit}.  A cross-check in which cubic polynomials were
fitted to each band produces peak magnitudes and dates of maximum in each band
consistent with the SALT2 answers, within the errors, for all of the new SNe;
we adopt the SALT2 values as our fiducial values for direct comparison with
other work.

We estimate the host reddening of the SNe in two separate ways.  First,
we fit the $B-V$ color behavior of each SN to the Lira relation
\citep{phillips99,csp10} for those SNe for which we have appropriate
light curve phase coverage.
The Lira relation is believed explicitly not to hold for SN~2007if and the
candidate super-Chandra SN~Ia~2009dc \citep{yamanaka09,taub11}, but its
value may nevertheless be useful in studying the relative intrinsic color of
these SNe.  Additionally, we search for \na\ absorption at the redshift of
the host galaxy for each SN.  We perform a $\chi^2$ fit to the \na\ line
profile, modeled as two separate Gaussian lines with full width at half
maximum equal to the SNIFS instrumental resolution of 6~\AA, to all SNIFS
spectra of each SN, as for SN~2007if in \citet{scalzo10}.  In the fit, the
equivalent width \ewna\ of the \na\ line is constrained to be non-negative.
We convert these to estimates of $E(B-V)_\mathrm{host}$ using both the
shallow-slope (0.16~\AA$^{-1}$) and steep-slope (0.51~\AA$^{-1}$) relations
from \cite{tbc02} (``TBC'').
While the precision of these relations has been called into question when
used on their own \citep[e.g.][]{poznanski11}, we believe that examining such
estimates together with the Lira relation and fitted colors from the light
curve can provide helpful constraints on the importance of host reddening.
The best-fit Lira excesses, values of \ewna, and derived constraints
on the host galaxy reddening are listed in Table~\ref{tbl:hostred}.

We detect weak \na\ absorption in SNF~20070803-005 and SNF~20080522-000.
Neither of these SNe appear to have very red colors according to the SALT2
fits, and we believe it to be unlikely that either are heavily extinguished,
so for purposes of extinction corrections to occur later in our analysis,
we use reddening estimates from the shallow-slope TBC relation, together
with a CCM dust law with $R_V = 3.1$ \citep{cardelli}.
When applied to Milky Way \na\ absorption in our spectra, the shallow-slope
TBC relation produces $E(B-V)$ estimates consistent with \cite{sfd}.
SNF~20070912-000 shows a marginal ($< 2\sigma$) detection, though the spectra
are noisy and the limits are not strong.  We detect no \na\ absorption in
the other SNe.

\begin{deluxetable}{lrrrr}
\tabletypesize{\footnotesize}
\tablecaption{Host galaxy reddening estimates from the Lira relation and
              \ewna\ fits \label{tbl:hostred}}
\tablehead{
   \colhead{} &
   \colhead{\ion{Na}{1}~D} &
   \colhead{Shallow} &
   \colhead{Steep} &
   \colhead{Lira} \\
   \colhead{SN Name} &
   \colhead{EW\tablenotemark{a}~(\AA)} &
   \colhead{TBC\tablenotemark{b}} &
   \colhead{TBC\tablenotemark{c}} &
   \colhead{Relation\tablenotemark{d}}
}
\startdata
SNF~20070528-003 & $0.00^{+0.32}_{-0.00}$ & $0.00^{+0.05}_{-0.00}$
                 & $0.00^{+0.12}_{-0.00}$ & \nodata \\[0.02in]
SNF~20070803-005 & $0.19^{+0.05}_{-0.06}$ & $0.02^{+0.01}_{-0.01}$
                 & $0.06^{+0.03}_{-0.03}$ & $0.06 \pm 0.06$ \\[0.02in]
SN~2007if        & $0.00^{+0.12}_{-0.00}$ & $0.00^{+0.01}_{-0.00}$
                 & $0.00^{+0.02}_{-0.00}$ & $0.13 \pm 0.07$ \\[0.02in]
SNF~20070912-000 & $0.32^{+0.44}_{-0.18}$ & $0.04^{+0.07}_{-0.03}$
                 & $0.12^{+0.22}_{-0.09}$ & \nodata \\[0.02in]
SNF~20080522-000 & $0.29^{+0.08}_{-0.08}$ & $0.04^{+0.01}_{-0.01}$
                 & $0.11^{+0.04}_{-0.04}$ & $0.13 \pm 0.12$ \\[0.02in]
SNF~20080723-012 & $0.00^{+0.19}_{-0.00}$ & $0.00^{+0.02}_{-0.00}$
                 & $0.00^{+0.06}_{-0.00}$ & $0.00 \pm 0.10$ \\
\vspace{-0.1in}
\enddata
\tablenotetext{a}{Measured from a simultaneous fit of the \na\
                  absorption line profile to all SNIFS spectra of each SN.}
\tablenotetext{b}{$E(B-V)$ derived from \na\ absorption, using the ``shallow''
                  slope (0.16~\AA$^{-1}$) of \citet{tbc02}.}
\tablenotetext{c}{$E(B-V)$ derived from \na\ absorption, using the ``steep''
                  slope (0.51~\AA$^{-1}$) of \citet{tbc02}.}
\tablenotetext{d}{Best-fit $E(B-V)$ from the Lira relation in the form given
                  in \citep{phillips99}.  Error bars include a 0.06~mag
                  intrinsic dispersion of normal SNe~Ia around the relation,
                  added in quadrature to the statistical errors.}
\end{deluxetable}

Based on the very strong limit on \na\ absorption from the host galaxy,
\citet{scalzo10} inferred that the large Lira excess of SN~2007if was not due
to host galaxy extinction.  Those SNe observed at sufficiently late phases
show measured Lira excesses consistent with zero, additional evidence that
host galaxy dust extinction is minimal for these SNe if the Lira relation
holds.

Allowing for varying amounts of extinction associated with the Lira excess
or \na\ absorption, each of the new SNe have maximum-light $(B-V)$ colors
consistent with zero.
While most of our SNe have well-sampled light curves around maximum light
and hence have well-measured maximum-light colors, SN~2007if has a gap
between $-9$~days and $+5$~days with respect to $B$-band maximum
(phases fixed by the SALT2 fit).
The light curve fit of \citep{scalzo10}, using SALT2 v2.0
and a spectrophotometric reduction using a previous version of the SNIFS
pipeline, suggests a red color $B-V = 0.16 \pm 0.06$.  The more recent
reduction fit with SALT2 v2.2 gives $B-V = 0.12 \pm 0.06$.
However, the directly measured $B-V$ color of SN~2007if at $-9$~days
($-0.07\pm 0.04$) and at $+5$~days ($0.15 \pm 0.02$) are each consistent
within the errors with the mean values at those epochs for our other five SNe.
The systematic error on the maximum-light color of SN~2007if, at least
0.04~mag, may therefore be too large for it to be considered significantly
redder at maximum than its counterparts.


\subsection{Bolometric Light Curve and \nickel\ Synthesis}
\label{subsec:bololc}

As input to our further analysis to calculate \nickel\ masses and total
ejected masses for our sample of SNe, we calculate quasi-bolometric UVOIR
light curves from the photometry.  To derive bolometric fluxes from SNIFS
spectrophotometry, we first deredden the spectra to account for Milky Way
dust reddening \citep{sfd}, then deredden by an additional factor
corresponding to a possible value of the host galaxy reddening, creating
a suite of spectra covering the range $0.00 < E(B-V)_\mathrm{host} < 0.40$
in 0.01~mag increments.
We then integrate the dereddened, deredshifted, flux-calibrated
spectra over all rest-frame wavelengths from 3200--9000~\AA.
For each quasi-simultaneous set of ANDICAM $BVRI$ observations and each
possible value of $E(B-V)_\mathrm{host}$, we multiply
the dereddened, deredshifted SNIFS spectrum nearest in time by a cubic
polynomial, fitted so that the synthetic photometry from the resulting
spectrum matches, in a least-squares sense, the ANDICAM imaging photometry
in each band.  We then integrate this ``warped'' spectrum to produce the
bolometric flux from ANDICAM.  This procedure creates a set of bolometric
light curves with different host galaxy reddening values which we can use
in our later analysis (see \S\ref{sec:modeling}).

SNF~20070528-003 and SNF~20070912-000 are at a higher redshift than our
other SNe, such that SNIFS covers only the rest-frame wavelength range
3000--8500~\AA, so we integrate their spectra in this range instead.
We expect minimal systematic error from the mismatch, since the phase coverage
for these two SNe is such that only the bolometric flux near maximum light,
when the SNe are still relatively blue, is useful for the modeling described
in \S\ref{sec:modeling}.

To account for the reprocessing of optical flux into the near-infrared (NIR)
by iron-peak elements over the evolution of the light curve, we must apply a
NIR correction to the integrated SNIFS fluxes.  No NIR data were taken for
any of the SNe except SN~2007if; in general they were too faint to observe
effectively with the NIR channel of ANDICAM.
Since these are peculiar SNe, any correction for the NIR flux necessarily
involves an extrapolation.  Since the $I$-band second maximum has low contrast
for all the SNe in our sample, the $JHK$ behavior should be similar for these
SNe to the extent that $I$ and $JHK$ are related
\citep[e.g. as in][]{kasen06}.  With these caveats, we therefore use the
NIR corrections for SN~2007if derived in \citet{scalzo10} for all of our SNe,
with the time axis stretched according to the
stretch factor derived from the SALT2 $x_1$ \citep{guy07} to account for the
different timescales for the development of line blanketing in these SNe.
Our modeling requires knowledge of the bolometric flux only near maximum light
(to constrain the \nickel\ mass) and more than 40 days after maximum light
(to constrain the total ejected mass).
The NIR correction is at a minimum near maximum light ($\sim 5\%$),
and at a maximum near phase $+40$~days ($\sim 25\%$), so it should not evolve
quickly at these times and our results should not be strongly affected.
We assign a systematic error of $\pm 5\%$ of the \emph{total} bolometric flux
(or about $\pm 30\%$ of the NIR flux itself near $+40$~d) to this correction.

To estimate \nickel\ masses, we also need a measurement of the bolometric
rise time $t_\mathrm{rise,bol}$.  We establish the time of bolometric maximum
light $t_\mathrm{max,bol}$ by fitting a cubic polynomial to the bolometric
fluxes in the phase range $-10~\mathrm{d} < t < +20~\mathrm{d}$.  We then
calculate the final bolometric rise time $t_\mathrm{rise,bol}$ via
\begin{equation}
t_\mathrm{rise,bol} = t_{\mathrm{rise,}B} - t_{\mathrm{max,}B}
   + t_{\mathrm{max,bol}}.
\label{eqn:trisebol}
\end{equation}
We find $t_\mathrm{max,bol}$ to occur about 1~day earlier than
$t_{\mathrm{max,}B}$ for the SNe in our sample.  We have a strong constraint
on the time of explosion only for SN~2007if \citep{scalzo10}, for which this
procedure results in a bolometric rise time of 23~days.  
We have utilized the discovery data from our search along with our spectrophotometry
to constrain the rise times of the new SNe presented here. For these we find 
$t_{rise,B}=21.2\pm1.9$~days, and so use $t_{rise,bol}=20\pm2$~days in our
models. Our value is very similar to the value of $t_{\mathrm{rise,}B}=21\pm2$~days
given by the sample of 1991T-like SNe~Ia in \citet{ganesh11}.
This approach leads to more
conservative uncertainties than the single-stretch correction of
\citet{conley06}; Figure~6 of \citet{ganesh11} suggests that the relation
between rise time and decline rate $\Delta m_{15}$ (or stretch $s$)
may break down at the high-stretch end.

We calculate the \nickel\ mass, $M_\mathrm{Ni}$, for the six SNe by relating
the maximum-light bolometric luminosity $L_\mathrm{bol}$ to the luminosity
from radioactive decay $L_\mathrm{rad}$ \citep{arnett82}:
\begin{eqnarray}
\alpha^{-1} L_\mathrm{bol} \ & = & \ L_\mathrm{rad} \nonumber \\
& = & \ N_\nickel \lambda_\nickel Q_{\nickel,\gamma} \,
   e^{-\lambda_\nickel t} \nonumber \\
& + & \ N_\nickel \lambda_\nickel
   \frac{\lambda_\cobalt}{\lambda_\nickel-\lambda_\cobalt}
      \left( Q_{\cobalt,e^+} + Q_{\cobalt,\gamma} \right) \nonumber \\
& & \ \ \ \ \times
     \left( e^{-\lambda_\cobalt t} - e^{-\lambda_\nickel t} \right),
\label{eqn:Lrad}
\end{eqnarray}
where $t$ ($= t_\mathrm{rise,bol}$ in this case) is the time since explosion,
$N_\nickel = M_\nickel/(56~\mathrm{AMU})$ is the number of \nickel\ atoms
produced in the explosion, $\lambda_\nickel$ and $\lambda_\cobalt$ are
the decay constants for \nickel\ and \cobalt\
($e$-folding lifetimes 8.8~days and 111.1~days) respectively,
and $Q_{\nickel,\gamma}$, $Q_{\cobalt,\gamma}$ and $Q_{\cobalt,e^+}$
are the energies released in the different stages of the decay chain
\citep{nad94}.  The dimensionless number $\alpha$ is a correction factor
accounting for the diffusion time delay of gamma-ray energy through the
ejecta, typically ranging between 0.8 and 1.6 for reasonable explosion models
\citep[see e.g., Table~2 of][where it is called $Q$]{hk96}.
A nominal value of $\alpha = 1.2$ is often used in the literature
\citep[e.g.][]{nugent95,bk95,howell06,howell09}.
The tamped detonation models of \citet{kmh93} and \citet{hk96},
on which we will base our modeling later in the paper, have slightly higher
values closer to $\alpha = 1.3$.
We therefore adopt a fiducial value of $\alpha = 1.3$ for our simple estimate
here.

\begin{deluxetable}{lrrr}
\tabletypesize{\footnotesize}
\tablecaption{\nickel\ mass reconstruction for the SNfactory SNe
              \label{tbl:56Ni}}
\tablehead{
   \colhead{SN Name} &
   \colhead{$L_\mathrm{bol}$ ($10^{43}$~erg~s$^{-1}$)} &
   \colhead{$t_\mathrm{rise,bol}$~(d)\tablenotemark{a}} &
   \colhead{$M_\mathrm{Ni}~(\Msol)$\tablenotemark{b}}
}
\startdata
SNF~20070528-003 & $1.87 \pm 0.06$ & $20 \pm 2$ & $0.78 \pm 0.11$ \\
SNF~20070803-005 & $1.62 \pm 0.05$ & $20 \pm 2$ & $0.69 \pm 0.07$ \\
SN~2007if        & $2.93 \pm 0.22$ & $23 \pm 2$ & $1.38 \pm 0.09$ \\
SNF~20070912-000 & $1.86 \pm 0.06$ & $20 \pm 2$ & $0.77 \pm 0.10$ \\
SNF~20080522-000 & $1.70 \pm 0.06$ & $20 \pm 2$ & $0.74 \pm 0.08$ \\
SNF~20080723-012 & $1.82 \pm 0.07$ & $20 \pm 2$ & $0.76 \pm 0.10$ \\
\vspace{-0.1in}
\enddata
\tablenotetext{a}{Calculated from Equation~\ref{eqn:trisebol}.}
\tablenotetext{b}{Assuming fiducial $\alpha = 1.3$.}
\end{deluxetable}

The resulting \nickel\ mass estimates are shown in Table~\ref{tbl:56Ni}.
The new SNe have $M_\mathrm{Ni}$ in the range 0.7--0.8~\Msol,
at the high end of what might be expected for Chandrasekhar-mass
explosions; the well-known W7 deflagration model \citep{w7} produced
0.6~\Msol\ of \nickel, while some delayed detonation models can produce
up to 0.8~\Msol\ \citep[e.g., the N21 model of][]{hk96}.


\subsection{Spectral Features and Velocity Evolution}
\label{subsec:specanal}

\begin{figure*}
\center
\resizebox{\textwidth}{!}{\includegraphics{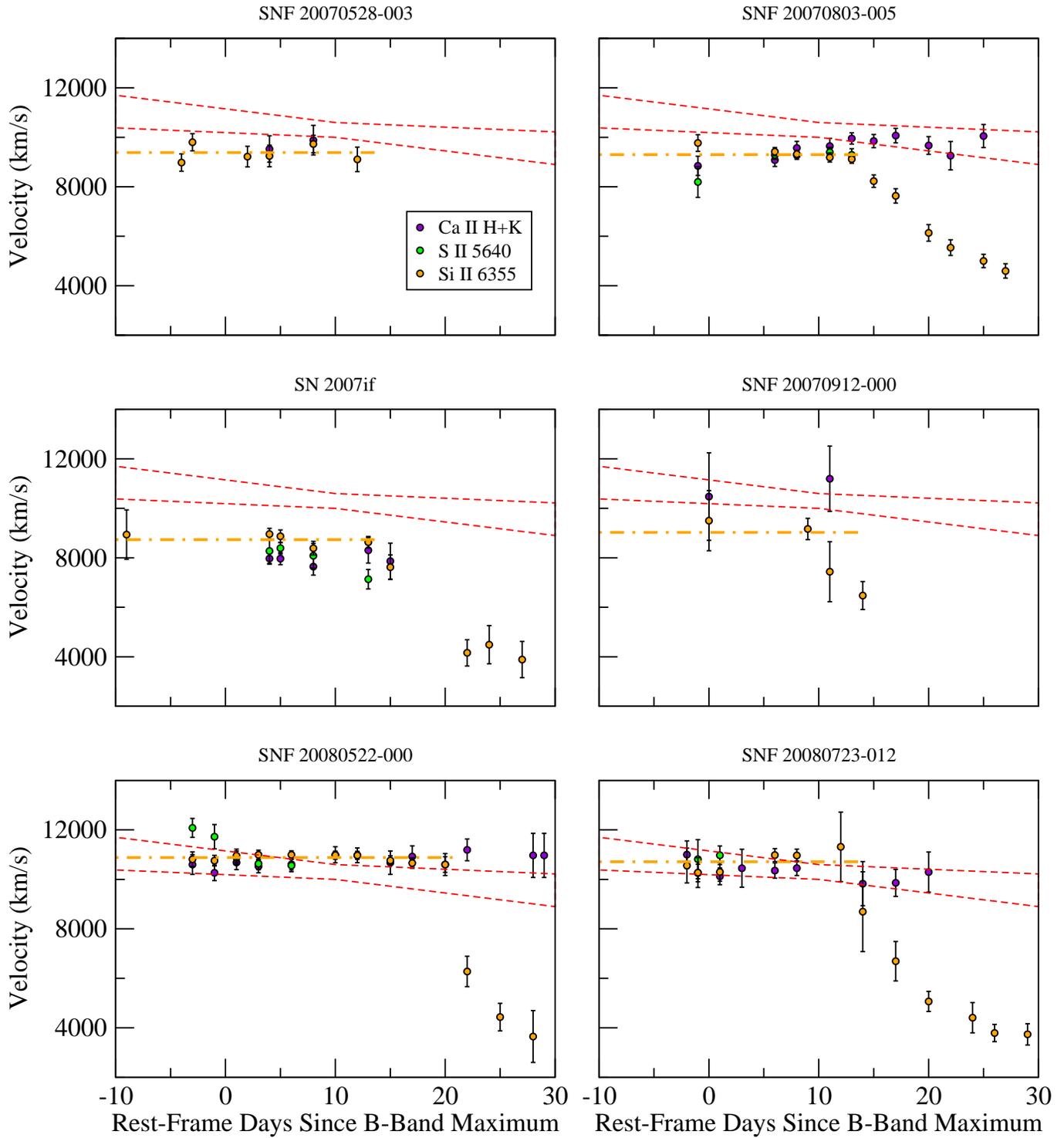}}
\caption{\small Time evolution of the velocities of various
intermediate-mass-element absorption minima from the
super-Chandra candidate sample:
\ion{Si}{2}~$\lambda 6355$, 
\ion{S}{2}~$\lambda 5640$, and \ion{Ca}{2}~H+K.  The thick dash-dot line
shows a $\chi^2$ fit to a constant for all \vSi\ data before the end of the
plateau phase for each SN.  The thin dashed lines show the 1-sigma range
of behavior for the LVG subclass of \citet{benetti05}.}
\label{fig:vlines}
\end{figure*}

\begin{figure}
\center
\resizebox{0.90\columnwidth}{!}
{\includegraphics{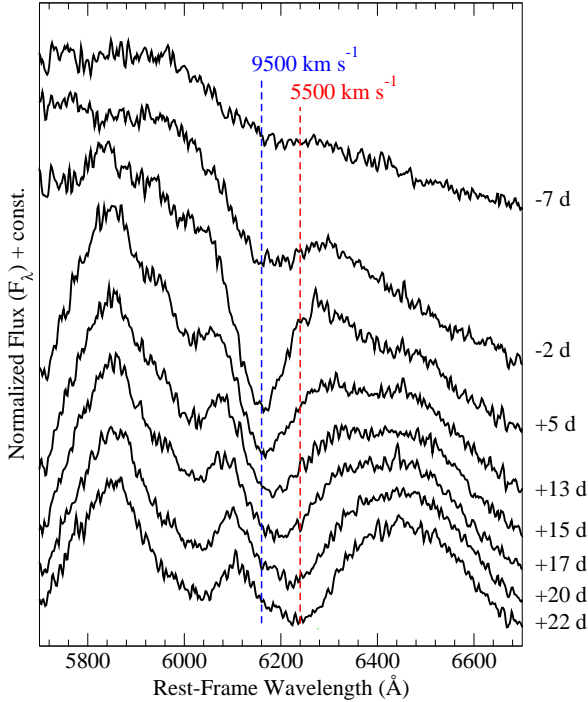}}
\caption{\small Detailed evolution of SNF~20070803-005 spectra around the
\ion{Si}{2}~$\lambda 6355$ feature.}
\label{fig:SiII-20070803005}
\end{figure}

The time evolution of the position of the \ion{Si}{2}~$\lambda 6355$
absorption minimum, showing the recession of the photosphere through the
ejecta, is shown in Figure~\ref{fig:vlines}.  The measurements of the
absorption minimum were made as follows:  bins in each spectrum immediately
to the right and left of the line feature were used to fit a linear
pseudocontinuum, $F_{\lambda,\mathrm{cont}} = a + b\lambda$.
This fitted pseudocontinuum was divided into each bin of the spectrum in the
region of the line.  The resulting spectrum was smoothed with a third-order
Savitzsky-Golay filter and the minimum was recorded as the bin with the
lowest signal.  The error bars on the procedure were determined through a
bootstrap Monte Carlo:  In the first stage, values of $a$ and $b$
representing possible pseudocontinua $F_{\lambda,\mathrm{cont}}$ were sampled
using the covariance matrix of the pseudocontinuum fit; for each candidate
pseudocontinuum, fluctuations typical of the measured errors on
each spectral bin were added to the smoothed spectrum, and the results were
smoothed and the minimum measured again.  We have verified that the
Savitzsky-Golay filter preserves the line minimum, so that smoothing a spectrum
twice introduces negligible systematic error.  The final velocity values and
their errors were measured as the mean and standard deviation of the
distribution of absorption minimum velocities thus generated.
While this method has slightly less statistical power than a fit to the
entire line profile as in \citet{scalzo10}, we believe it is more robust to
possible bias from line profiles with unusual shapes, and provides more
realistic error bars for the line minimum.



\begin{deluxetable}{lr@{\,$\pm$\,}rr@{\,$\pm$\,}rrrr}
\tabletypesize{\footnotesize}
\tablecaption{Velocity gradient characteristics\label{tbl:vSi}}
\tablehead{
   \colhead{SN Name} &
   \multicolumn{2}{c}{$v_\mathrm{pl}$\tablenotemark{a}} & 
   \multicolumn{2}{c}{$\dot{v}$\tablenotemark{b}} &
   \colhead{$t_\mathrm{pl,0}$\tablenotemark{c}} & 
   \colhead{$\Delta t_\mathrm{pl}$\tablenotemark{d}} &
   \colhead{$\chi^2_\nu$\tablenotemark{e}}
}
\startdata
SNF~20070528-003 &  9371 & 171 & $ -6$ &  29 & $-4$ & $16$ & 0.85 \\
SNF~20070803-005 &  9695 &  81 & $ 46$ &  23 & $-1$ & $14$ & 1.00 \\
SN~2007if        &  8963 & 248 & $ 31$ &  29 & $-9$ & $21$ & 0.76 \\
SNF~20070912-000 &  9201 & 403 & \multicolumn{2}{c}{\nodata}
                 & \nodata & \nodata & \nodata \\
SNF~20080522-000 & 10936 & 107 & $  5$ &  10 & $-4$ & $16$ & 0.41 \\
SNF~20080723-012 & 10391 & 291 & $-72$ &  47 & $-2$ & $10$ & 0.72 \\
\vspace{-0.1in}
\enddata
\tablenotetext{a}{Best-fit constant velocity, in \kms, with error.}
\tablenotetext{b}{Slope of the best-fit line to absorption line velocities
                  from the first reliable measurement until the break
                  associated with \ion{Fe}{2} line blending,
                  in \kms~day$^{-1}$.}
\tablenotetext{c}{Phase of first available measurement, in days with respect
                  to $B$-band maximum light, which we interpret
                  to be the start of the plateau.}
\tablenotetext{d}{Minimum duration of the visible plateau phase in days,
                  until \ion{Si}{2} becomes blended with \ion{Fe}{2}.}
\tablenotetext{e}{Chi-square per degree of freedom for fit to a constant.}
\end{deluxetable}

We found that for lines with equivalent widths less than 15~\AA,
the absorption minima had unreasonably large uncertainties and/or showed large
systematic deviations from the trend described by stronger measurements.
When measuring these absorption minima, we are probably simply measuring
uncertainty in the pseudocontinuum.  \citet{blondin11} saw similar effects
when measuring velocities of very weak absorption minima, to the extent that
the \ion{Si}{2}~$\lambda 6355$ velocity would even be seen to increase with
time.  We therefore reject measurements of such weak absorption features.

Our candidate super-Chandra SNe~Ia share a slow evolution of the \ion{Si}{2}
velocity, consistent within the errors with being constant in time for each
SN from the earliest phases for which measurements are available.
Table \ref{tbl:vSi} shows the fitted constant velocities and the chi-square
per degree of freedom, $\chi^2_\nu$, for a fit to a constant.
For comparison with earlier work, the velocity gradient $\dot{v} = -dv/dt$
calculated as the slope of the best-fit linear trend of the measurements
before day $+14$, is also listed, along with the formal error from the fit.
All of our SNe would be classified as Benetti LVG \citep{benetti05}
based on their velocity gradients.  While the slope of the straight-line fit
to the absorption velocities for SNF~20070803-005 seems to differ from zero
at the $2\sigma$ level (formal errors), the reduced chi-square for this fit
is extremely small ($\chi^2_\nu = 0.006$) and we conclude that the evolution
cannot be reliably distinguished from a constant ($\chi^2_\nu = 1.00$).
Similarly, the best-fit line to the absorption velocities for SNF~20080723-012
has a positive slope at $1.5\sigma$ ($\dot{v} = 72 \pm 47~\kms$~day$^{-1}$),
but once again, a constant is a good fit to the data ($\chi^2_\nu = 0.72$).

The slow \ion{Si}{2} absorption velocity evolution usually lasts until about
two weeks after maximum light, after which a break in the behavior occurs and
a pronounced decline begins,
at a rate of \mbox{$\sim 500~\kms~\mathrm{day}^{-1}$}.  At this point,
developing \ion{Fe}{2} lines have probably blended with \ion{Si}{2} and made
the velocity measurements unreliable \citep[see, e.g.][]{phillips92}.
The transition occurs concurrently with the onset of the second maximum in
the $I$-band light curve, which may be attributed to light reprocessed from
bluer wavelengths by the recombination of \ion{Fe}{3} to \ion{Fe}{2} as
the ejecta expand and cool \citep{kasen06}.
Figure~\ref{fig:SiII-20070803005} shows the evolution of the
\ion{Si}{2}~$\lambda 6355$
line profile for SNF~20070803-005 near the transition.  The line profile
after the break shows a portion of the \ion{Si}{2} line near the plateau
velocity, suggesting that the material responsible for the plateau has
thinned but is not yet transparent.
We mark the phases of the last \vSi\ measurement visually consistent with the
plateau velocity, and the first measurement inconsistent with it,
by vertical dashed lines in Figure~\ref{fig:bvrilc} to show their
correspondence with the $I$-band second maximum.

In four of the six SNe, the plateau velocity is low ($\sim~9000$~km~s$^{-1}$),
inconsistent with the normal range of behavior of the LVG subclass of
\citet{benetti05}.  The remaining two SNe, SNF~20080522-000 and
SNF~20080723-012, have higher plateau velocities ($\sim 11000$~km~s$^{-1}$),
falling roughly into the range of LVG behavior, although their velocity
gradients are still flatter than any remarked upon in that work.
Due to the lower S/N of our spectra of SNF~20070912-000 and the weakness of
the \ion{Si}{2}~$\lambda 6355$ absorption feature, only two measurements of
the line velocity before the break, each around 9000 km/s with large errors,
were extracted; we can at most say that the observed velocity gradient in
this SN is consistent with that of the other SNe in our sample.

Figure~\ref{fig:vlines} also shows the line minimum velocities of
\ion{Ca}{2}~H+K and \ion{S}{2}~$\lambda\lambda 5456, 5640$.
Although, as expected, \ion{Ca}{2}~H+K stays optically thick longer
than \ion{Si}{2}~$\lambda 6355$, its velocities at early times are consistent
with the observed plateau behavior shown by \ion{Si}{2}.  The \ion{S}{2} lines
are weak, but when their absorption minima can be reliably measured, they do
not show dramatically higher or lower velocities than the other lines.
This supports the interpretation that all of these lines are formed in the
same thin, dense layer of ejecta.


\section{Constraints on Total Mass and Density Structure}
\label{sec:modeling}

\newcommand{\fFe}{\ensuremath{f_\mathrm{Fe}}}
\newcommand{\fNi}{\ensuremath{f_\mathrm{Ni}}}
\newcommand{\fSi}{\ensuremath{f_\mathrm{Si}}}
\newcommand{\fCO}{\ensuremath{f_\mathrm{CO}}}
\newcommand{\fsh}{\ensuremath{f_\mathrm{sh}}}
\newcommand{\fenv}{\ensuremath{f_\mathrm{env}}}
\newcommand{\MWD}{\ensuremath{M_\mathrm{WD}}}
\newcommand{\Msh}{\ensuremath{M_\mathrm{sh}}}
\newcommand{\Menv}{\ensuremath{M_\mathrm{env}}}
\newcommand{\Mtot}{\ensuremath{M_\mathrm{tot}}}
\newcommand{\vsh}{\ensuremath{v_\mathrm{sh}}}
\newcommand{\ve}{\ensuremath{v_e}}

The modeling procedure we use here represents a refined version of that
used in \citet{scalzo10}, which we compare and contrast with the similar
approach of \citet{stritz06} in \S\ref{subsec:exprho}
for the special case of a simple equivalent exponential
density profile.  We then describe our extensions to the method, including
modeling of density profiles with shells (\S\ref{subsec:exprhosh}),
priors on the central density (\S\ref{subsec:pc}), 
and calculation of the \nickel\ form factor $q$ (\S\ref{subsec:qint}).
We then present our final modeling results in \S\ref{subsec:model-results}.


\subsection{SN~Ia Ejected Mass Measurements Using
            the Equivalent Exponential Density Formalism}
\label{subsec:exprho}

The ejecta density structure of SNe~Ia is frequently modeled as an exponential
$\rho(v) \propto \exp(-v/v_e)$, where the ejecta are in homologous expansion
at velocity $v = (r-r_\mathrm{initial})/t$ since the explosion at time
$t = 0$, and $v_e$ is a characteristic velocity scale.  Many hydrodynamic
models of SN~Ia explosions, including the well-known W7 model \citep{w7},
have density profiles which are very close to exponential.

\citet{jeffery99} made semianalytic calculations of the time evolution of
the gamma-ray energy deposition in an exponential model SN~Ia, with reference
to its bolometric light curve.  By about 60 days after explosion, virtually
all of the \nickel\ has decayed and the dominant energy source is the decay
of \cobalt\ (which in turn was produced by \nickel\ decay at earlier times).
The optical depth to Compton scattering of \cobalt\ gamma rays behaves as
$\tau_\gamma = (t/t_0)^{-2}$ with $\tau_\gamma = 1$ at some fiducial
time $t_0$.  The value of $t_0$ can be extracted by fitting the bolometric
light curve for $t > 60$~d to a modified version of Equation~\ref{eqn:Lrad},
in which $\alpha = 1$ (rather than its maximum-light value from Arnett's rule)
and the term corresponding to \cobalt\ gamma rays is multiplied by a factor
$1-\exp(-\tau_\gamma)$.  The total mass of the ejecta can then be expressed as
\begin{equation}
\MWD = \frac{8\pi}{\kappa_\gamma q} (v_e t_0)^2.
\label{eqn:MWD}
\end{equation}
Here, $\kappa_\gamma$ is the Compton scattering opacity
for \cobalt\ gamma rays,
and $q$ is a form factor describing the distribution of the \cobalt\ in the
ejecta, which follows the original distribution of \nickel\ in the explosion.
The value of $\kappa_\gamma$ is expected to lie in the range
0.025--0.033~cm$^{-2}$~g \citep{swartz95}, with the low end (0.025)
corresponding to the optically thin regime.  The value of $q$ can be
readily calculated given an assumed distribution of \nickel\
(see \S\ref{subsec:qint} below).

\citet{stritz06} used this method to measure progenitor masses
for a sample of well-observed SNe~Ia with $UBVRI$ light curve coverage.
They constructed ``quasi-bolometric'' light curves according to the procedure
of \citet{contardo00}, by converting the observed $UBVRI$ magnitudes to
monochromatic fluxes at the central wavelengths of their respective filters,
then summing them, using corrections for lost flux between filters derived
from spectroscopy of SN~1992A.  They then applied the semianalytic approach
of \citet{jeffery99} to fit for the gamma-ray escape fraction, and hence
the ejected mass of the SN~Ia progenitor.

Our own work improves on previous use of this method in two important ways.
First, \citet{stritz06} made no attempt to correct for the NIR contribution
to the bolometric flux,
simply asserting that it is small during the epochs of interest.
In \citet{scalzo10} we found that for SN~2007if the NIR contribution
was indeed small ($\sim 5\%$) near maximum light, but was greater than 25\%
at 40~days after maximum light, and our estimate for its value at 100~days
after maximum light is still around 10\%.  Therefore, at least for SNe~Ia
like the ones we study here, the method of \citet{stritz06} underestimates
the fraction of trapped \cobalt\ gamma-rays for a given initial \nickel\ mass,
and hence the ejected mass, as a result of neglecting NIR flux.

Second, our fitting procedure includes covariances between different inputs
to the prediction for the bolometric light curve, constrained by a set of
Bayesian priors motivated by explosion physics.  Specifically, covariances
between $q$, $v_e$, and $\alpha$ may influence the
interpretation of the fitted value of $t_0$.
\citet{stritz06} simply fixed the \nickel\ mass from Arnett's rule, and then
fit for $t_0$.  Similarly, they assume $q = 1/3$, $v_e = 3000$~km~s$^{-1}$,
and $\alpha = 1$ for all of their SNe, with no covariance between any
of these parameters.
Because models with more \nickel\ need less gamma-ray trapping to produce
the same bolometric luminosity at a given time, there is a large fitting
covariance between the \nickel\ mass and $t_0$, mentioned in \cite{scalzo10}.
The value of $\alpha$ is model-dependent, and not
a fundamental physical quantity, but as noted in \S\ref{subsec:bololc}
above, $\alpha = 1.2$ ($\pm 0.1$) is also a common choice
when no other prior is available from explosion models.
Since $\alpha$ affects the nickel mass, a smaller assumed value of $\alpha$
results in a larger \nickel\ mass, but a smaller \emph{ejected} mass,
as interpreted from a given bolometric light curve.
In a self-consistent choice of parameters, $v_e$ and $q$
will each depend in part on the mass of \nickel\ and therefore on $\alpha$.

The value of $v_e$ is difficult to measure directly, since observed velocities
of absorption line minima may depend on temperature as much as density.
Since $v_e$ appears squared in Equation~\ref{eqn:MWD}, its contribution to the
error budget on \MWD\ is potentially quite large if treated as an independent
input.  However, its value can be constrained
within a range of $\pm 300$~km~s$^{-1}$ by requiring energy conservation.
Following practice in the literature \citep{howell06,maeda09b}, we calculate
the kinetic energy $E_K$ as the difference between the energy $E_N$ released
in nuclear burning and the gravitational binding energy $E_G$, and then
set $v_e = (E_K/6\MWD)^{1/2}$.

Calculating the energy budget of a SN~Ia requires us to assume a composition.
Our model considers four components to the ejecta:
\begin{itemize}
\item \nickel, which contributes to the luminosity, $E_N$, and $E_G$;
\item Stable Fe-peak elements (``Fe''), which contribute to $E_N$ and $E_G$;
\item Intermediate-mass elements such as Mg, Si and S (``Si''),
      which contribute to $E_N$ and $E_G$;
\item Unburned carbon and oxygen (``C/O''), which contribute only to $E_G$.
\end{itemize}
The input parameters, which we vary using a Metropolis-Hastings Monte Carlo
Markov chain, are the white dwarf mass \MWD, the central density $\rho_c$
(needed in the calculation of $E_G$), the parameter $\alpha$ from Arnett's
rule, the bolometric rise time $t_\mathrm{rise,bol}$, and the fractions
\fNi, \fFe, and \fSi\ of \nickel, stable Fe, and intermediate-mass elements
within \MWD.  We fix the fraction of unburned carbon and oxygen
$\fCO = 1-\fFe-\fNi-\fSi$.  We use the prescription of \citet{maeda09b}
to determine $E_N$:
\begin{equation}
E_N = \left[1.74 \fFe + 1.56 \fNi + 1.24 \fSi\right]
      \left( \frac{\MWD}{\Msol} \right) \times 10^{51} \mathrm{erg}.
\label{eqn:EN}
\end{equation}
We use the binding energy formulae of \citet{yl05} for
$E_G = E_G(\MWD,\rho_c)$, where $\rho_c$ is the white dwarf central density.
These ingredients determine $v_e$.  We apply Gaussian priors
$\alpha = 1.3 \pm 0.1$ \citep[as for SN~2007if;][]{scalzo10} and
$t_\mathrm{rise,bol} = 20 \pm 2$~days, and on $E(B-V)_\mathrm{host}$
according to the ``shallow TBC'' values in Table~\ref{tbl:hostred}.
We adopt $\kappa_\gamma = 0.025$~cm$^{-2}$~g
after \citet{jeffery99} and \citet{stritz06}.

One limitation with our approach is the use of $E_G$ from \citet{yl05},
derived for supermassive, differentially rotating white dwarfs.  This formula
remains an easily accessible estimate in the literature for the binding energy
of a white dwarf over a wide range of masses, used by several other authors
\citep{howell06,jbb06,maeda09b}.  The models of \citet{yl05} have been
criticized on the grounds that they may not exist in nature \citep{piro08},
nor explode to produce SNe~Ia if they do exist \citep{sn04,pfannes10a}.
However, it seems reasonable to assume that such models could represent
a snapshot in time of a rapidly rotating configuration, such as that
encountered in a white dwarf merger, which then detonates promptly rather
than continuing to exist as a stable object.  The merger simulations of
\citet{pakmor11} and \citet{pakmor12}, though they produce comparatively
little \nickel, show that prompt detonations in violent mergers can occur.
\citet{pfannes10b} simulated prompt detonations of rapidly rotating white
dwarfs with masses up to 2.1~\Msol, and found that the amount of \nickel\
produced could be as high as 1.8~\Msol, similar to SN~2007if.

In summary, our procedure directly extracts from a fit to the bolometric
light curve only the quantities $M_\mathrm{Ni}$ and $t_0$
(Equation~\ref{eqn:Lrad}), and marginalizes, in effect,
over $\alpha$, $t_\mathrm{rise,bol}$, $v_e$, and $q$.  The front end of our
modeling technique varies the mass, composition, and structure of the SN
progenitor as physical quantities which we wish to constrain.
We then convert these inputs into physically motivated priors on the values
of $v_e$ and $q$, using Equations~\ref{eqn:Lrad}, \ref{eqn:MWD},
and \ref{eqn:EN}, and finally calculate the ejected mass \MWD.


\subsection{Including the Effects of a Shell}
\label{subsec:exprhosh}

The above considerations all apply to conventional exponential-equivalent
models of expanding SN~Ia ejecta.  To explain the velocity plateaus of the
SNe in our sample, however, our model has a disturbed density structure where
the high-velocity ejecta
(included in the mass \MWD\ which undergoes nuclear burning)
are compressed into a dense shell of mass \Msh, traveling at velocity \vsh.
In tamped detonation models, such as the explosion models DET2ENV2, DET2ENV4
and DET2ENV6 \citep[][hereafter ``DET2ENVN'']{kmh93,hk96}, such a shell is
formed at the reverse shock of the interaction of the ejecta of an otherwise
normal SN~Ia with a compact ($\sim 10^{10}$~cm) envelope of material
with mass \Menv\ (\emph{external} to, and \emph{not} included in, \MWD).
The suffix $N$ in DET2ENVN refers to the envelope mass,
so for example model DET2ENV2 has $\Menv = 0.2~\Msol$.
``Pulsating delayed detonation'' models, such as the PDD535 model of
\citet{hk96}, have similar shells created by non-homologous pulsations of the
white dwarf progenitor prior to the final explosion, and hence do not require
an external shell of material.  However, these models tend
to produce fainter events, with much shorter rise times and redder colors,
than we observe for our sample, and so we do not consider them here.

In a tamped detonation, the material which will form the shell imparts its
momentum to the envelope, which in the DET2ENVN models acquires an average
velocity of about $1.5\vsh$.  The interaction ends within about the first
minute after explosion, and the shell then expands homologously with the
other ejecta thereafter.  We observe \vsh\ directly as the plateau velocity,
allowing us to constrain \Msh\ and \Menv, and, indirectly, the kinetic energy
scale $v_e$ of the ejecta.  For a given value of $v_e$ and a measured value
of \vsh, and neglecting the binding energy of the envelope,
conservation of momentum gives \citep[for more detail see][]{scalzo10}
\begin{equation}
\Menv = \frac{2}{3} \left[ \frac{3 v_e}{\vsh}
     Q\left(4,\frac{\vsh}{v_e}\right)
   - Q\left(3,\frac{\vsh}{v_e}\right) \right] \MWD.
\label{eqn:vsh}
\end{equation}
where $Q(a,x) = \gamma (a,x) / \Gamma(a)$ is the incomplete gamma function.
We calculate $\fenv = \Menv/\MWD$ and $\fsh = \Msh/\MWD$
by solving Equation~\ref{eqn:vsh} numerically.  In a double-degenerate merger
scenario, the total system mass $\Mtot = \MWD + \Menv$ is then equal to the
initial mass of the two white dwarfs undergoing the merger.

For these calculations, we use only the \emph{velocity} of the
\vSi\ plateau measured from our spectroscopic time series.  We do not model
the \emph{duration} of the plateau or the behavior of \vSi\ after the plateau
phase ends.  While the detailed evolution of \vSi\ undoubtedly contains useful
information, reproducing it would require detailed calculations of synthetic
spectra which are beyond the scope of this paper.  However, as long as we
have enough measurements of \vSi\ to show that a given SN exhibits plateau
behavior, we can reliably measure $\vsh$ without knowing the opacity of the
material in the shell.  We will compare our observations to previous numerical
models and observations of SNe~Ia with interacting shells in
\S\ref{subsec:shell-comp}.



\subsection{Including the Effects of Central Density on $Fe$ Yields}
\label{subsec:pc}

In \citet{scalzo10}, the stable iron fraction \fFe\ was allowed
to vary freely.  In this situation, \fFe\ and \fSi\ are nearly degenerate,
since the contribution per unit mass of Fe to the nuclear energy released
is only about 40\% higher than that of Si.  However, since Fe is produced
by neutronization in the densest parts of the ejecta during the explosion,
a high value of \fFe\ can greatly reduce $q$ because the formation of a
large Fe core displaces \nickel\ to a higher average velocity and a lower
optical depth.  It therefore becomes important to constrain Fe production
in any model in which we attempt to calculate $q$.

The DET2ENVN explosion models, on which our models are loosely based,
were intended to describe the detonation of a low-density white dwarf merger
remnant of mass 1.2~\Msol\ inside envelopes of varying mass.
The central density $\rho_c$ in these models is $4 \times 10^7$~g~cm$^{-3}$,
substantially lower than typical central densities of $\sim 10^9$~g~cm$^{-3}$
of deflagrations and delayed detonation models in the literature
\citep{w7,hk96,krueger10}.  These models have no stable Fe cores
immediately after explosion, with $X_\nickel = 0.9$ throughout the region
where \nickel\ and Fe are produced.

\citet{krueger10} investigated the effects of central density on \nickel\
yields in 3-D simulations of detonations of Chandrasekhar-mass white dwarfs,
averaging over an ensemble of realizations for each value of $\rho_c$.
They found $\fNi/(\fNi+\fFe) = 0.9$ for $\rho_c = 10^9$~g~cm$^{-3}$,
decreasing by $0.047$ on average for each $10^9$~g~cm$^{-3}$ increase
of $\rho_c$ thereafter.

Our model already includes the effects of the central density $\rho_c$
on the binding energy $E_G$ of the white dwarf, through the fitting formula
of \citet{yl05}.  Although \MWD\ may be super-Chandrasekhar in our models,
the overall extent of nuclear burning should depend on the density,
not the mass, and so we may consider extrapolating those results here.
Since the link between \fFe\ and \fNi\ is statistical rather than
deterministic, we do not attempt to calculate a definite fraction for each
model.  Instead, we calculate the ratio of \nickel\ to total iron-peak
elements, $\eta = \fNi/(\fNi+\fFe)$, as well as the relation from
\citet{krueger10}:
\begin{equation}
\eta_\mathrm{Kr10} = \min \left\{ 0.9, 0.959 - 0.047
      \left( \frac{\rho_c}{10^9~\mathrm{g~cm}^{-3}} \right) \right\}
\end{equation}
enforcing a Gaussian prior $\eta = \eta_\mathrm{Kr10} \pm 0.1$.
For most models calculated, this results in only a small fraction of
stable Fe, as appropriate for low-density explosions.


\subsection{Calculation of $q$ for SN~Ia Models with a Shell}
\label{subsec:qint}

The gamma-ray transport form factor $q$ is the dimensionless \nickel-weighted
gamma-ray optical depth through the ejecta.
Its value ranges between 0 and 1, with large values corresponding to
high central concentrations of \nickel; the gamma-ray optical depth is
proportional to $q$.  For perfectly mixed, exponentially distributed
ejecta where the fraction of \nickel\ is constant throughout, $q = 1/3$.

\begin{figure}
\center
\resizebox{\columnwidth}{!}{\includegraphics{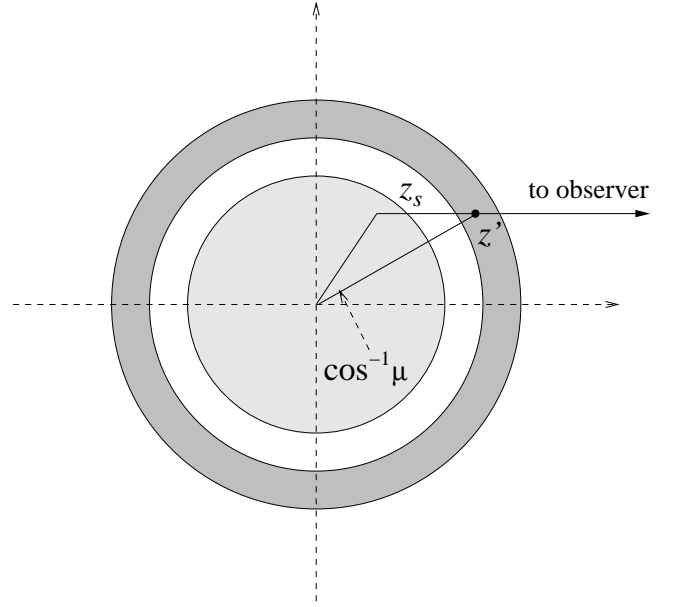}}
\caption{\small Integration geometry for the \nickel\ distribution form factor
$q$.  A \cobalt\ gamma-ray is emitted from point $z$ in the region where
$X_\nickel(z) > 0$ (inner shaded region), and travels along the ray $z_s$
towards the observer.  Point $z'$ lies in a mass shell of thickness $dz'$
(shaded annulus) in which the gamma-ray is trapped, scatters and deposits
its energy.}
\label{fig:qgeom}
\end{figure}

In \citet{scalzo10}, the calculation of the total mass for SN~2007if
assumes $q = 1/3$.  We chose this value because the lack of a distinct
second maximum in the $I$-band light curve suggested that a large amount of
\nickel\ had to be mixed to higher velocities
(see \S\ref{subsec:reddening} above).
This is not necessarily true for our other SNe.  The use of $q = 1/3$ also
assumes that the reverse-shock shell has a negligible effect on gamma-ray
trapping, which may also not be true for very massive shells.  Fortunately,
$q$ is easy to calculate numerically \citep{jeffery99}:
\begin{equation}
q = \frac{\int_0^\infty dz \int_0^\infty dz_s \int_{-1}^1 d\mu \
          z^2 \tilde\rho(z) X_\nickel(z) \tilde\rho(z')}
         {\int_0^\infty dz \ z^2 \tilde\rho(z)  X_\nickel(z)},
\end{equation}
where $z$ and $z'$ are dimensionless velocity coordinates in units of $v_e$,
$z_s = \sqrt{z'^2+z^2+2zz'\mu}$ is the beam path length,
$\tilde\rho(z)$ is a dimensionless density profile normalized to unit
mass, and $X_\nickel(z)$ is the (velocity-dependent) \nickel\ fraction just
after explosion.  The geometry of the integration is shown in
Figure~\ref{fig:qgeom}.

\begin{figure}
\center
\resizebox{\columnwidth}{!}{\includegraphics{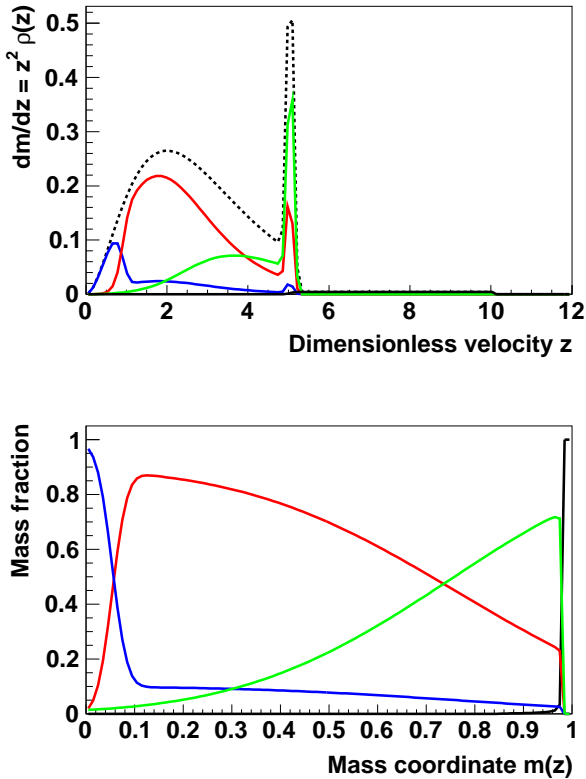}}
\caption{\small Density and composition for an example model with
$\fFe = 0.12$, $\fNi = 0.6$, $\fsh = 0.12$, $\fenv = 0.02$.  Top:  Mass of
a shell of ejecta of thickness $dz$ as a function of $z$.  Black dotted line:
overall density; solid lines, in rough order of increasing $z$:
stable Fe-peak (blue), \nickel\ (red), IME + unburned C/O (green),
C/O envelope (black).  Bottom:  Composition of model as a function of
mass coordinate $m(z)$.}
\label{fig:density-comp}
\end{figure}

\begin{deluxetable*}{lrrrrrrrr}
\tabletypesize{\footnotesize}
\tablecaption{Extracted model parameters}
\tablehead{
   \colhead{SN Name} &
   \colhead{$M_\mathrm{tot}/\Msol~(>98\%)$\tablenotemark{a}} &
   \colhead{$M_\mathrm{WD}/\Msol~(>98\%)$\tablenotemark{b}} &
   \colhead{$M_\mathrm{Ni}/\Msol$\tablenotemark{c}} &
   \colhead{$f_\mathrm{env}$\tablenotemark{d}} &
   \colhead{$f_\mathrm{sh}$\tablenotemark{e}} &
   \colhead{$f_\mathrm{Fe/sh}~(<98\%)$\tablenotemark{f}} &
   \colhead{$\chi^2_\nu$\tablenotemark{g}} &
   \colhead{$P_\mathrm{fit}$\tablenotemark{h}}
}
\startdata
SNF~20070528-003 & \phantom{$2.04^{+0.47}_{-0.39}$}~$(>1.49)$
   & \phantom{$1.89^{+0.48}_{-0.41}$}~$(>1.32)$
   & $0.91 \pm 0.12$ & $0.08 \pm 0.05$
   & $0.30 \pm 0.12$ & $0.23^{+0.24}_{-0.14}~(<0.66)$ & $1.08$ & $0.290$
   \\[0.02in]
SNF~20070803-005 & $1.57^{+0.18}_{-0.16}~(>1.32)$
   & $1.38^{+0.15}_{-0.13}~(>1.18)$ & $0.75 \pm 0.09$ & $0.14 \pm 0.03$
   & $0.40 \pm 0.04$ & $0.40^{+0.16}_{-0.12}~(<0.68)$ & $0.69$ & $0.654$
   \\[0.02in]
SN~2007if        & $2.30^{+0.27}_{-0.24}~(>1.94)$
   & $1.98^{+0.21}_{-0.18}~(>1.70)$ & $1.37 \pm 0.13$ & $0.16 \pm 0.04$
   & $0.44 \pm 0.05$ & $0.63^{+0.17}_{-0.18}~(<0.85)$ & $1.32$ & $0.077$
   \\[0.02in]
SNF~20070912-000 & \phantom{$2.05^{+0.45}_{-0.39}$}~$(>1.50)$
   & \phantom{$1.90^{+0.46}_{-0.41}$}~$(>1.33)$
   & $0.89 \pm 0.15$ & $0.09 \pm 0.05$
   & $0.31 \pm 0.12$ & $0.22^{+0.25}_{-0.13}~(<0.66)$ & $0.10$ & $0.775$
   \\[0.02in]
SNF~20080522-000 & $1.56^{+0.19}_{-0.14}~(>1.34)$
   & $1.45^{+0.17}_{-0.13}~(>1.25)$ & $0.80 \pm 0.11$ & $0.08 \pm 0.02$
   & $0.28 \pm 0.04$ & $0.33^{+0.15}_{-0.11}~(<0.59)$ & $0.37$ & $0.976$
   \\[0.02in]
SNF~20080723-012 & $1.79^{+0.28}_{-0.21}~(>1.49)$
   & $1.69^{+0.25}_{-0.18}~(>1.41)$ & $0.84 \pm 0.11$ & $0.06 \pm 0.02$
   & $0.25 \pm 0.05$ & $0.24^{+0.12}_{-0.08}~(<0.45)$ & $1.09$ & $0.324$
   \\[0.02in]
\tableline
SN~1991T       & $1.65^{+0.22}_{-0.16}~(>1.39)$
   & $1.50^{+0.18}_{-0.13}~(>1.28)$ & $0.77 \pm 0.15$ & $0.10 \pm 0.03$
   & $0.34 \pm 0.05$ & $0.30^{+0.16}_{-0.11}~(<0.60)$ & $0.83$ & $0.600$
   \\[0.02in]
SN~2003fg      & \phantom{$2.31^{+0.30}_{-0.36}$}~$(>1.77)$
   & \phantom{$2.01^{+0.31}_{-0.36}$}~$(>1.46)$
   & $1.18 \pm 0.16$ & $0.16 \pm 0.07$
   & $0.44 \pm 0.10$ & $0.47^{+0.24}_{-0.21}~(<0.83)$ & $0.09$ & $0.945$ \\
\vspace{-0.1in}
\enddata
\label{tbl:confint}
\tablecomments{Quantities with error bars are marginalized over all
independent parameters.  Uncertainties are 68\% CL ($1\sigma$) and represent
projections of the multi-dimensional PDF onto the derived quantities.
Upper or lower limits on poorly constrained properties are 98\% CL.}
\tablenotetext{a}{Total system mass of the two merged white dwarfs.}
\tablenotetext{b}{Mass of the central white dwarf merger product which
                  undergoes nuclear burning in the explosion.}
\tablenotetext{c}{\nickel\ mass synthesized in the explosion.}
\tablenotetext{d}{Ratio of envelope mass to central merger product mass.}
\tablenotetext{e}{Fraction of mass of burnt ejecta which is compressed into
                  the reverse-shock shell.}
\tablenotetext{f}{Mass fraction of iron-peak elements (\nickel\ + stable Fe)
                  in the reverse-shock shell.}
\tablenotetext{g}{Minimum $\chi^2_\nu$ achieved by a fit to the data.}
\tablenotetext{h}{Probability of attaining the given value of $\chi^2_\nu$
                  or higher if the model is a good fit to the data,
                  incorporating all priors.}
\end{deluxetable*}

To calculate $q$, we assume a density profile motivated by DET2ENVN which
includes an envelope with density profile $\rho \sim z^{-2}$ on the outside,
a thin Gaussian shell at $z_\mathrm{sh} = \vsh/v_e$, and an exponential
in velocity beneath the shell.  The total density is given by
\begin{eqnarray}
\rho(z) & = & C_\mathrm{bulk} \phi\left(\frac{z-z_\mathrm{sh}}
              {\sigma_\mathrm{z,sh}}\right) \exp(-z)
          +   C_\mathrm{env} \phi\left(\frac{z_\mathrm{sh}-z}
              {\sigma_\mathrm{z,sh}}\right) z^{-2} \nonumber \\
        & + & C_\mathrm{sh} \exp \left[ -\frac{1}{2} \left(
              \frac{z-z_\mathrm{sh}}{\sigma_\mathrm{z,sh}}\right)^2 \right],
\end{eqnarray}
where we choose $\phi(z) = \erfc(z/\sqrt{2})/2$ as a smooth function
which approaches 1 for $z \ll -1$ and 0 for $z \gg +1$, and the constants
$C_\mathrm{bulk}$, $C_\mathrm{env}$, $C_\mathrm{sh}$, are determined
so that the mass fractions in each density profile component agree with
the input values.  \citet{qhw07} suggest a nominal width of 500~km~s$^{-1}$
for the reverse-shock shell at velocity $\vsh \sim 10000$~km~s$^{-1}$.
The self-similar shock interaction model of \citet{chevalier82} suggests that
the reverse shock velocity width should be about 3\% of the velocity at the
contact discontinuity for an interaction with a $r^{-2}$ envelope and ejecta
with a power-law density profile $r^{-n}$ with $n \sim 7$
\citep[used to approximate SNe~Ia in the context of interaction
       with a CSM wind, e.g.,][]{mwv04}.
We assume a shell half-width of $\sigma_\mathrm{z,sh} = 0.015 z_\mathrm{sh}$
in line with \citet{chevalier82}, although the expansion is homologous and
no longer self-similar after the interaction.  Our results are not sensitive
to the exact value of $\sigma_\mathrm{z,sh}$;
values in the range (0.01--0.05)$z_\mathrm{sh}$ give us the same
value of $q$ to within 5\% for reverse-shock shells with masses up to
$0.5\MWD$, and with much better agreement for less massive shells.

For $X_\nickel(z)$, we use a parametrized composition structure inspired
by \citet{kasen06}:
\begin{eqnarray}
X_\nickel(z) & = & \phi\left(\frac{m_\mathrm{Fe,core}-m(z)}
                      {a_\mathrm{Fe,core}}\right) \nonumber \\
        & \times & \phi\left(\frac{m(z)-m_\mathrm{IPE}}
                      {a_\mathrm{IPE}}\right) \times \max(\eta,0.9),
\end{eqnarray}
where $m(z) = \int_{0}^{z} z^2 \tilde\rho(z) \, dz$ is the mass coordinate,
and the stable Fe-peak core and the \nickel\ mixing zone are bounded by mass
coordinates $m_\mathrm{Fe,core}$ and $m_\mathrm{IPE}$ with mixing widths
$a_\mathrm{Fe,core}$ and $a_\mathrm{IPE}$, respectively.  This allows for some
stable Fe-peak elements to be mixed throughout while maintaining a core in the
innermost regions for high central density models
(see Figure~\ref{fig:density-comp}).  For our modeling below
we choose $a_\mathrm{Fe} = 0.025$ and $a_\mathrm{Ni} = 0.35$, so that the
outward mixing of \nickel\ corresponds roughly to the ``enhanced mixing''
case of \citet{kasen06}, in accordance with the behavior we see in the
light curves.
As the mass of the shell increases, the results may depend more
sensitively on the particular distribution of \nickel\ in the shell.
We proceed with the analysis, but caution that more detailed models of the
shock interaction, and/or actual hydrodynamic simulations of the explosion,
may be needed to accurately understand gamma-ray transport for cases in
which a large amount of \nickel\ is swept up into the shell.

In general, the values of $q$ are higher for our SNe ($0.45 \pm 0.05$)
than the nominal $q = 1/3$ value for a completely mixed exponential SN~Ia,
but there is very little variation with shell mass fraction.
The mixing of \nickel\ into a shell (and potentially above the photosphere)
and the displacement of \nickel\ to higher velocities by a stable Fe core
have comparable effects on $q$, but the former effect is minimized in
the ``enhanced mixing'' model characteristic of our light curves.


\subsection{Modeling Results}
\label{subsec:model-results}

\begin{figure*}
\center
\resizebox{0.9\textwidth}{!}{\includegraphics{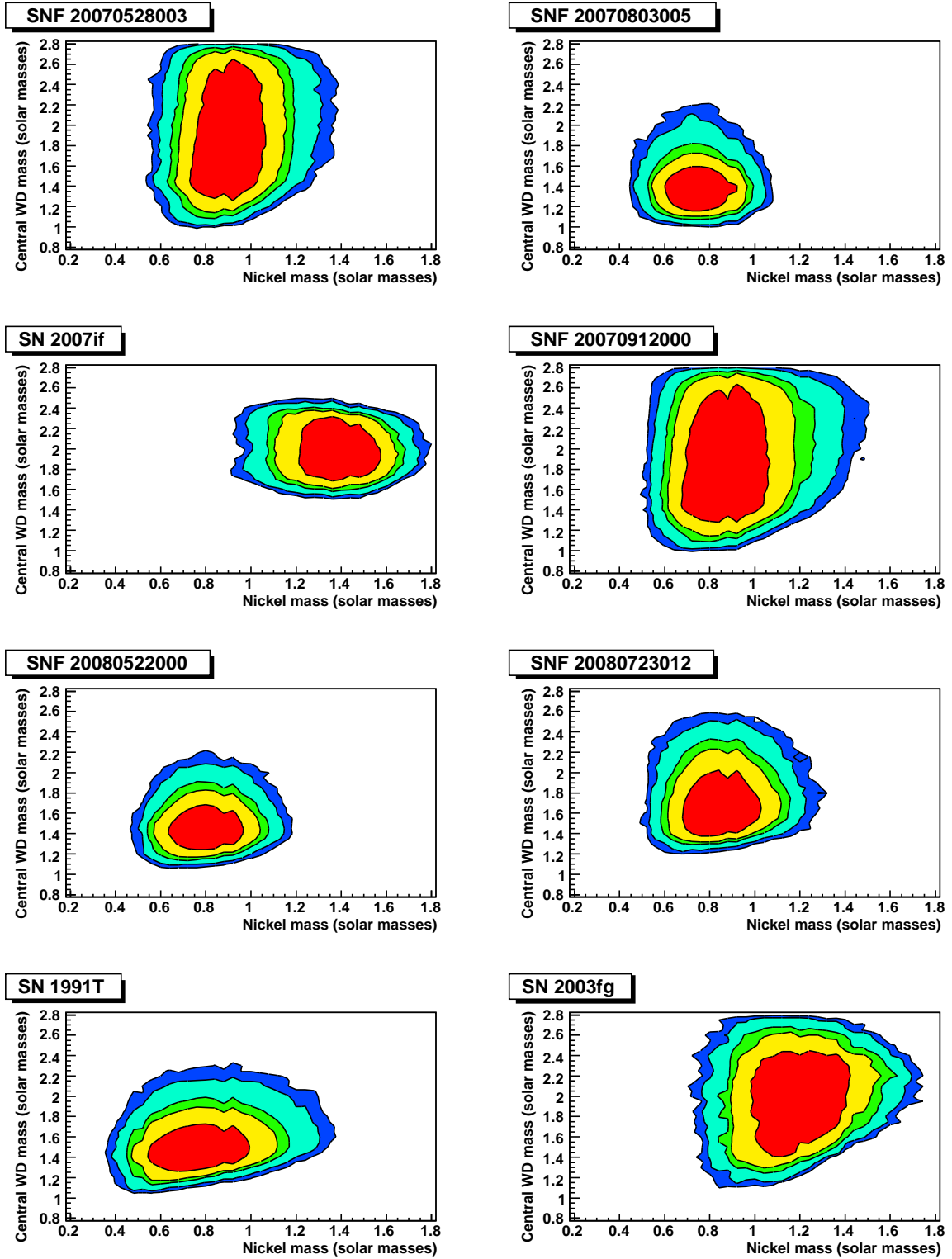}}
\caption{\small Confidence regions in the \MWD-\nickel\ mass plane for
our six SNe, with priors $\alpha = 1.3 \pm 0.1$ and $\eta = 0.90 \pm 0.05$
as for DET2ENVN.  SN~1991T and SN~2003fg are also shown.
Contours bound regions of constant probability density.
Colored regions are 68\% (red), 90\%, 95\%, 99\%, and 99.7\% CL (blue).}
\label{fig:mtot_all}
\end{figure*}

The results of our modeling for the SNe in our sample are summarized
in Table~\ref{tbl:confint}.  Figure~\ref{fig:mtot_all} shows the confidence
regions in the \MWD-\nickel\ mass plane for all six SNfactory SNe.
While we limit our later statements about relative rates
(see \S\ref{subsec:rates}) and Hubble residuals (see \S\ref{subsec:hubble})
to the untargeted SNfactory search, which samples the smooth Hubble flow
($z > 0.03$) and samples all host galaxy environments in an unbiased manner,
we also include our results for two spectroscopically analogous SNe~Ia from
the literature as useful points of comparison:  SN~1991T and SN~2003fg
(see Appendix~\ref{sec:litsne}).

Our results can be summarized as follows for the SNe in the SNfactory sample:
\begin{itemize}
\item We recover the results of \citet{scalzo10} for SN~2007if, within
      the uncertainties.  The total system mass has come down slightly,
      from 2.41~\Msol\ to 2.30~\Msol\ (probability distribution median),
      since trapping of \cobalt\ gamma-rays by the envelope is now
      included in the mass estimate.  The fractions of the total system
      mass in the shell and in the envelope, set by the plateau velocity,
      remain the same.
\item SNF~20070803-005 and SNF~20080522-000 are consistent with being
      Chandrasekhar-mass SNe~Ia, albeit spectroscopically peculiar ones.
\item SNF~20070528-003 and SNF~20070912-000 have data quality sufficient
      only to obtain lower limits on the total mass, and upper limits on
      the shell and envelope mass.  In particular the lack of late-time
      photometry points mean that the lower limits are driven by the \nickel\
      mass from Arnett's rule.  The total system mass \Mtot, which includes
      the mass \Menv\ of the envelope which we infer from the plateau
      velocity, is super-Chandrasekhar-mass at $> 98\%$ confidence,
      although the mass \MWD\ of the central merger remnant is consistent
      with Chandrasekhar-mass values.
\item SNF~20080723-012 appears to be a moderately super-Chandrasekhar-mass
      object, with \nickel\ mass, total mass, and other parameters
      intermediate between SN~2007if and the rest of the population.
      We place a 98\% CL lower limit of 1.41~\Msol\ on \MWD, so that the
      progenitor system is likely super-Chandrasekhar-mass even without
      including \Menv.
\end{itemize}
We also perform the following cross-checks.  First, our results are not
strongly sensitive to the assumption of a particular mixing parameter,
changing by less than 2\% if we instead assume a completely stratified
composition with $a_\mathrm{Fe} = a_\mathrm{Ni} = 0$.  Although changes in
the mixing parameters do influence $q$ substantially for models with small
amounts of \nickel, they matter much less when the \nickel\ mass is large.
We also confirm that when we discard the velocity information and fix
$\Menv = 0$, the resulting median reconstructed masses \Mtot\ are within 2\%
of the median values of \MWD, and the probability distributions have
comparable widths.  This is expected, since only the most massive envelopes
(as in SN~2007if) make an appreciable contribution to the gamma-ray optical
depth as seen from the inner layers of ejecta.


\section{Discussion}
\label{sec:discussion}


We have shown that the subset of SNfactory events selected based on initial
spectra similar to SN~2003fg, SN~2007if and SN~1991T can be fit well by a
simple semi-analytic model of a tamped detonation, intended to describe
the results of mergers of double-degenerate systems with total mass
at or exceeding the Chandrasekhar limit.  In the following section
we explore results in the literature which can help us determine
the extent to which this interpretation is unique and relevant.

In \S\ref{subsec:shell-comp}, we compare our observations to numerical models
of SNe~Ia with interacting shells in their ejecta, and to observations of
SN~2005hj taken by \citet{qhw07}.
In \S\ref{subsec:fryer}, we evaluate the possibility that some of the
luminosity of our events may be due to an ongoing shock interaction.
In \S\ref{subsec:maeda}, we discuss other recent models of asymmetric,
single-degenerate SN~Ia explosions which attempt to reproduce the velocity
plateau phenomenon, and their implications both for whether our SNe are
super-Chandrasekhar and whether they are mergers.
In \S\ref{subsec:rates}, we present the relative rates and examine the extent
to which our results may constrain merger scenarios if a significant number
of SNe~Ia have double-degenerate progenitors.
In \S\ref{subsec:ruiter}, we compare the aggregate probability distribution
of total system mass from our events to population synthesis models
of super-Chandrasekhar-mass double-degenerate mergers.
Finally, in \S\ref{subsec:hubble}, we examine the deviation from the Hubble
diagram for these SNe and discuss implications for SN~Ia cosmology.


\subsection{Comparison with Shell SN~Ia Models in the Literature}
\label{subsec:shell-comp}

Our most recent point of comparison in the literature for the application of
explosion models with interacting shells to observations of SNe~Ia is
\citet{qhw07}.  They noted the presence of a velocity plateau in their
observations of SN~2005hj and compared them to delayed detonation,
pulsating delayed detonation and tamped detonation models
\citep{kmh93,hk96,gerardy04}.

For a shell mass fraction $\fsh = 0.07$, similar to our less extreme SNe,
the predicted $B-V$ color is around 0.05-0.1.  \citet{qhw07} note that the
systematic uncertainty in the absolute value of $(B-V)_\mathrm{max}$ may be
as large as 0.1 mag for the DET2ENVN models and other shell models.
The color they report for SN~2005hj is 
$(B-V)_\mathrm{max} = 0.04 \pm 0.06$, consistent with the mean color in
our sample, for a claimed shell mass fraction $\fsh = 0.14$, comparable to
SNF~20070803-005.

According to \citet{qhw07} and references therein, one might expect to see
cooler photospheres, and hence redder $(B-V)_\mathrm{max}$, with increasing
shell mass fraction.
Our modeling predicts that SN~2007if, with the lowest plateau velocity and
the most massive progenitor, has the most massive shell in both absolute
and relative terms.  This SN may be somewhat redder than the others near
maximum light, but the lack of light curve coverage near maximum makes it
difficult to say exactly how much.
Table~\ref{tbl:confint} also shows that SN~2007if has the largest expected
fraction of iron-peak elements in its shell, corresponding to the material
near the photosphere around maximum light.  To the extent that SN~2007if is
intrinsically redder than our other SNe, this may be due in part to line
blanketing by iron-peak elements, rather than a low-temperature photosphere,
which would be inconsistent with the weakness of \ion{Si}{2} $\lambda 5800$
in all the SNe in our sample.

While we cannot predict from theory how long the plateau phase should last
without more sophisticated modeling of our SNe, the durations of our plateaus
are also broadly consistent with expectations.  For models of
Chandrasekhar-mass events with envelope masses about 0.1~\Msol \citep{qhw07},
the plateau phase is expected to last about 10~days.  In most of our SNe
it lasts at least 15~days (and at least 10 days for SNF~20070912-000).
\citet{qhw07} derive a plateau duration of $20 \pm 10$ days for SN~2005hj.

SNF~20080522-000 shows the longest-lived plateau
in our data set, stretching as early as 10 days before $B$-band maximum and
lasting as long as 30 days.  A more conservative estimate would be that
the plateau phase is confirmed to last from day $-3$,
when \ion{Si}{2}~$\lambda 6355$ becomes strong enough that the error bars
on the velocity of the absorption minimum drop below 500~km~s$^{-1}$,
to day $+15$, where \ion{Fe}{2} lines begin to develop near
\ion{Si}{2}~$\lambda 6355$ and blending may become a concern.  All of our
velocity measurements in this 18-day time window are contained in a narrow
range just 250~km~s$^{-1}$ wide.


\subsection{Constraints on Ongoing Extended Emission}
\label{subsec:fryer}

\citet{fryer10} ran three-dimensional smoothed-particle hydrodynamics (SPH)
simulations of double-degenerate mergers.  They found that the central merger
remnants are indeed surrounded by an envelope
with an approximate radial density profile $\rho(r) \propto r^{-\beta}$
with $3 < \beta < 4$.  They then performed radiation hydrodynamics
simulations of the interaction of the SN ejecta with the envelope, calculating
synthetic spectra and light curves of the resulting explosions.
For sufficiently massive envelopes (more massive than about 0.1~\Msol),
energy advected by the shock is released over the evolution of the SN,
producing non-negligible luminosity near maximum light and extending
emission into UV wavelengths.
Based on these findings, \citet{fryer10} argued that these ``enshrouded''
systems would, in all likelihood, look nothing like SNe~Ia.
\citet{bs10} performed analogous calculations using the \texttt{STELLA}
radiation hydrodynamics code, finding that in general a radial density
profile as steep as $r^{-4}$ looked similar to a normal SN~Ia in optical
wavelengths ($UBVRI$), whereas a profile varying as $r^{-3}$ would be
dominated by shock emission.

These findings put significant constraints on the radial extent and density
profile of any envelope which might have enshrouded the progenitors of the
SNe in our sample.  In particular, the blue, mostly featureless spectrum
seen in SN~2007if at phase $-9$~days is characteristic of what \citet{fryer10}
expect, and the optical emission could be powered in part by advected heat
energy from the shock interaction at early times, resulting in a broader
light curve.  However, the fact that \vSi\ in our events is seen to assume
its plateau value as early as a week before maximum light, and the colors
appear similar to those of normal SNe~Ia near maximum light, argue strongly
against any ongoing interaction with an extended envelope or wind.

As noted in \citet{scalzo10},
if some fraction of the maximum-light luminosity is due to shock heating,
whether advected or resulting from a fresh interaction, instead of
\nickel\ decay, our mass estimates would tend to increase.  This is because
the influence of the shock interaction would diminish substantially
by about 50 days after explosion \citep{bs10}, and more \cobalt\ gamma rays
would have to be trapped in order to reproduce the observed light curve.
Any model in which shock heating produces a large amount of luminosity
more than 50 days after explosion would probably not look like a SN~Ia,
and could not explain the appearance of the SNe~Ia discussed in this paper.

One straightforward way to address the extent of any ongoing consequences of
shock interactions in future studies would be to obtain early-phase UV light
curves of a candidate super-Chandra SN~Ia from a satellite such as
\emph{Swift}.  The signature of any strong influence of shock heating would
be evident therein.


\subsection{Implications of Possible Asymmetry}
\label{subsec:maeda}

The large inferred masses of our candidate super-Chandra SNe~Ia,
taken together with the observed plateaus in the \ion{Si}{2} velocity,
are both naturally explained by the tamped detonation model we put forth
in this paper, and by the underlying double-degenerate merger scenario
it represents.  Some recent work, however, points towards the possibility
of explaining these events in terms of asymmetric single-degenerate
explosions.

\citet{maeda09a,maeda10a,maeda10b,maeda11} invoke large-angular-scale
asymmetries to explain the diversity of velocity gradients observed in normal
SN~Ia explosions and the low/high velocity gradient (LVG/HVG) dichotomy
\citep{benetti05}.  They suggest that an asymmetric explosion may cause an
overdensity in the ejecta on one side of the explosion, causing LVG behavior
when viewed from that side, with the other side relatively less dense and
exhibiting HVG behavior.  Additionally, \citet{hachisu11} suggested that
optically-thick winds blown from an accreting white dwarf could strip mass
from the outer layers of its donor star, regulating the accretion rate and
potentially allowing the white dwarf to accrete without exploding until
reaching masses as large as 2.7~\Msol.  Such a white dwarf would have to
rotate differentially, as in the models of \citet{yl05}, with the attending
uncertainty in the evolutionary history of such objects.  
If a differentially rotating white dwarf were to explode asymmetrically,
this could present an explanation for our observations within the
single-degenerate scenario.  A scenario like this one, if correct, could
also explain SN~2007if and the HVG~SN~2009dc as being similar objects
viewed from different angles.  \citet{tanaka10} interpreted the low continuum
polarization of SN~2009dc as evidence for a nearly spherical explosion,
but low polarization could also be observed in an axisymmetric explosion
viewed along the symmetry axis, as those authors note.

While we cannot at this time conclusively rule out the possibility
that the \ion{Si}{2} velocity plateaus we observe in our SNe result from
asymmetry, no asymmetry is \emph{needed} as yet to explain them,
as argued e.g. by \citet{maeda09b} for super-Chandrasekhar-mass explosions.
The physical cause of the plateau --- overdensity in the ejecta ---
is the same in symmetric and asymmetric models.
If the velocity plateaus are indeed the result of asymmetric explosions,
the shell mass fractions derived in Table~\ref{tbl:confint} would still
have meaning in terms of disturbances to the density structure along the
line of sight, but the inferred envelopes would not be present, i.e.,
\Menv\ would be zero and \Mtot\ would equal \MWD\ for the SNe analyzed
in this paper.

We can use the relative rate of supernovae spectroscopically similar to
those in our sample (see \S\ref{subsec:rates} below) to make some general
statements about how well they can be explained by lopsided asymmetric
explosions.  If the SNe~Ia in our sample belonged to the same population as
normal SNe~Ia, and the spectroscopic peculiarity and brightness were due
entirely to viewing angle effects
\citep[see][for an asymmetric model for which this is true]{kasen04},
a relative rate of $\sim 2\%$ would imply a range of viewing angles no more
than $\sim 15$~degrees from the symmetry axis.  For models in which the
asymmetry does not translate into a very peculiar spectrum, as seems likely
for the \citet{maeda10b,maeda11} models with high \nickel\ mass, we should look
instead at the diversity of velocity gradients among spectroscopic analogues.
Since \emph{all} of the (spectroscopically similar) SNe~Ia in our sample have
velocity gradients at the slowly-evolving extreme of what the model of
\citet{maeda10b} claims to produce, it seems plausible that, within the
context of this class of models, the density enhancements in the ejecta
of our SNe are roughly isotropic.

Asymmetries resulting in overdensities in SN~Ia ejecta along the line of
sight could also affect gamma-ray trapping for that line of sight only.
Most of the trapping happens in low-velocity ejecta
(see \S\ref{subsec:model-results}), so the form factor $q$ should not be
strongly affected; we think it unlikely that $q$, and our ejected mass
estimates, could vary by more than 10\%.  Our \nickel\ mass estimates from
Arnett's rule should also be robust, since according to \citet{maeda11},
the effects of asymmetry on peak brightness are least pronounced for the most
luminous SNe~Ia.  Based on these considerations and those of the preceding
paragraph, the super-Chandrasekhar-mass status of SN~2007if and
SNF~20080723-012 seems secure to us.

A different class of asymmetries might arise in bipolar, rather than lopsided,
explosions.  An axially symmetric merger geometry, featuring a disk of
disrupted white dwarf material in the equatorial plane, could in principle
result in velocity plateaus and peculiar spectra when viewed edge-on, but a
more normal SN~Ia appearance when viewed pole-on.  This possibility was
recently suggested by \citet{lp11}, who also suggested that SNe~Ia from
merger events might show increasing diversity in their velocity gradients
with increasing mass of carbon.  In this scenario, SN~2009dc could be an
event like SN~2007if viewed pole-on, consistent with its low continuum
polarization \citep{tanaka10}; SN~2007if should then correspond to the edge-on
case, and should show substantial polarization.  \citet{lp11} make no
predictions for relative rates in different SN~Ia subgroups, which may depend
on details of the hydrodynamic interaction of the SN ejecta with the disk.
Nevertheless, an occluded region of ejecta narrow enough to explain their rare
incidence 
would subtend only a small solid
angle at the source, with only a small fraction of the ejecta decelerated.
This would lead to broader, more complex line profiles than the ones we
observe.  It therefore once again seems reasonable that our SNe should be
more or less spherically symmetric, with only moderate large-angular-scale
asymmetries.  Furthermore, any asymmetries compatible with our observations
should influence only the inferred envelope masses \Menv, and should leave our
estimates of \MWD\ intact at the 10\% level, as we noted above.

Future observations, including nebular spectra of candidate
super-Chandra SNe~Ia for direct comparison with the abovementioned
studies, may allow us to make more specific statements about their
asphericity.  Spectropolarimetry, as for SN~2009dc \citep{tanaka10},
may also be helpful.


\subsection{Constraints on Merger Scenarios from Relative Rates}
\label{subsec:rates}


Mass modeling shows that all six of our SNfactory-discovered SNe~Ia
have a high probability of having super-Chandrasekhar-mass progenitors.
Because ours was a wide-field search and our spectroscopy screening
was conducted impartially, that is without pre-selection based on
colors, luminosity, lightcurve shape, host galaxy environment etc.,
we may estimate the relative fraction of such objects.

A total of 400 spectroscopically-confirmed SNe~Ia were discovered
by the SNfactory. We did not spectroscopically screen candidates
found to be fainter on the discovery image than on any pre-discovery
detections that may have existed, as such candidates were assumed
to be after maximum light. Furthermore, we did not perform spectroscopic
screening in the cases where a host galaxy had a previously-known
redshift beyond $z=0.08$, though such cases were rare.

We consider two subsamples --- one that is purely flux-limited
and another that is further limited to $z\le0.08$. For both
subsamples we consider only SNe~Ia discovered at or before maximum
light, when the characteristic spectral features of SN~2003fg,
SN~2007if and SN~1991T are clearest \citep{li11a}.  The flux-limit
is represented by the survey efficiency as a function of magnitude
rather than as a fixed magnitude.  A subset redshift-limited to $z \leq 0.08$
is also considered as a cross-check, since here details of the survey
efficiency become unimportant and the spectroscopic typing is very
secure.

%

Among the super-Chandra candidates, all survive the pre-maximum and 
flux-limit cuts, while SNF~20070528-003 and SNF~20070912-000 are eliminated
by the redshift cut. The sample cuts applied to the overall sample leave
240 SNe~Ia in the flux-limited sample and 141
in the redshift-limited sample.  Due to their enhanced brightness and
generally longer rise times, the super-Chandra candidates are enhanced by
factors of $1.30\pm0.06$ and $1.11\pm0.01$ in each of the two subsamples.
These enhancements are calculated using the control time in the SNfactory 
search, accounting for both apparent magnitudes and light-curve widths of 
these SNe relative to normal SNe~Ia.  The assigned uncertainties are 
illustrative of the effect of a generous $\pm0.25$~mag shift in the 
detection efficiency curve. As expected, this affect is small and has 
essentially no impact for the volume-limited subset.  Including these 
correction factors, the rates are $1.9^{+1.1}_{-0.5}$\% and 
$2.6^{+1.9}_{-0.7}$\% for the flux-limited and redshift-limited 
subsamples, respectively. The difference in these relative rates is 
consistent with Poisson fluctuations, including accounting for the SNe~Ia 
in common.

These are the relative rates under the assumption that all of our 
candidates have super-Chandrasekhar-mass progenitors, as their estimated 
masses under the tamped detonation model indicate.  Alternatively, these 
can be taken as the rates for SNe~Ia classified as like SN~1991T by using 
SNID. Although our 1.9--2.6\% relative rate is somewhat lower than the 
``volume-limited'' rate of $9.4^{+5.9}_{-4.7}$\% (5-day cadence) from 
\cite{li11a}, we note that it is consistent with the $\sim 1\%$ ``Ia-91T'' 
rate of \citet{silverman12} from the same search. The difference between 
the two is related to how the events are classified. \citet{li11a} use 
classifications from the IAU Circulars, rather than a homogeneous set of 
spectra over a given range of wavelengths and light-curve phases 
subclassified with a single method; in fact, \citet{silverman12} imply 
that some of the \citet{li11a} subclassifications may be photometric 
(based on light-curve width) rather than spectroscopic. \citet{li11a} also 
treat SNe with type ``Ia-99aa'' as being ``Ia-91T' rather than breaking 
them into separate subclasses, which led to their much higher rate. Using 
SNID, \citet{silverman12} classify only one of the \citet{li11a} 
1991T-likes (SN~2004bv) as ``Ia-91T'' and the rest as ``Ia-99aa'' or 
normal, leading to an updated ``volume-limited'' rate of 
$1.4^{+3.0}_{-0.0}$ for the \citet{li11a} sample, consistent with our 
rate. As a point of comparison from an untargeted search similar to 
SNfactory's, rather than a search targeting known galaxies, we also 
consider the SDSS spectroscopic sample \citep{ostman11}. Seventy-eight of 
the 141 SNe~Ia in the sample of \citet{ostman11} have spectra with 
light-curve phases at or before maximum light, proof against the 
1991T-like ``age bias'' \citep{li11a}. Using SNID, \citet{ostman11} type 
only two of these 78 as ``Ia-91T'', leading to a relative rate of 
$3.4^{+2.4}_{-1.7}$\%, again consistent with our own rate. We note that 
our rate estimates have considerably smaller uncertainty than these other 
1991T-like SNe~Ia rates. The combination of all three surveys, using our 
redshift-limited rates, gives a net 1991T-like rate of 
$2.3^{+1.2}_{-0.6}$\%.


The statistical confidence intervals from our models indicate the possibility
that not all candidates are super-Chandrasekhar-mass. The probability density
extending below the Chandrasekhar mass represents the equivalent
of 0.2 SNe~Ia for both subsamples. If we correct for this excess
probability, the rates are lowered by factors of 5\% and 3\% for the
flux-limited and redshift-limited subsamples. A more precise
correction would depend on the true parent mass distribution function,
but it is evident from the above calculation that such details are
well below the Poisson uncertainties.

Finally, we consider the rates under the possibility that the shell
model is not appropriate. In this case there would be no contribution
from the shell mass, and the best mass estimate would be that of the
central merger remnant, \MWD. In this circumstance SNF~20070803-005
and SNF~20080522-000 could be considered more likely to have been
Chandrasekhar mass events.  The rates resulting from their removal
would then become $1.5^{+1.1}_{-0.4}$\% and $1.0^{+1.4}_{-0.4}$\%.
While we view this situation as unlikely, for reasons discussed above in
\S\ref{subsec:maeda}, the rates remain dominated by Poisson uncertainties
rather than modeling assumptions.

%

These relative rates are very low.
Therefore, if double-degenerate mergers are to contribute significantly
to the total SN~Ia rate, most such mergers would need to escape our
spectroscopic selection. Moreover, the SNfactory has not found overluminous
SNe~Ia in addition to those presented here or SN~2005gj \citep{snf2005gj}
that would qualify for our redshift-limited sample.

While our events are overluminous with large \nickel\ masses,
theoretical mergers \citep[e.g.]{pakmor11,pakmor12} thus far predict a range
of smaller \nickel\ masses, resulting in fainter explosions
with lower photospheric temperatures.
As discussed in \S\ref{subsec:fryer} above, a delay in the release of radiant
energy from a shock interaction could also contribute to heating
the photosphere at early times in events like SN~2007if, so that mergers with
less massive envelopes would have more normal spectra.

Merger events with less massive envelopes should also have less disturbed
density structures, and hence may not display the low \ion{Si}{2} velocity
plateau behavior we observe in our sample.
The small envelope mass fractions implied for the majority of merger events
would then represent a constraint on either the dynamics of the merger in
the case of a prompt explosion, or the post-merger evolution of the system
in the case of a delayed explosion triggered by accretion onto, or by
post-accretion spin-down of, the central merger remnant.
Unambiguous detection of a density enhancement signature for less massive
shells would be difficult. Construction of their mass distribution function
would require a parent sample of SNe~Ia all followed spectroscopically
such that any merger candidates could be selected 
by their velocity plateau behavior rather than spectroscopic peculiarity.
Moreover, such a study would require spectra at very early phases
($-10$~days or younger), since a less massive reverse-shock shell would
become transparent at earlier phases, and might not subsume most of the
intermediate-mass elements in the explosion.

Determining the relative importance of any of these factors, and thus
understanding the true relation of candidate super-Chandra SNe~Ia to the
general SN~Ia population, will require further work and new spectroscopic
data sets.
However, we can say that a range of envelope masses are probably needed
to explain the range of plateau velocities we observe in our data.


\subsection{Progenitor Mass Distribution
            in the Double-Degenerate Scenario}
\label{subsec:ruiter}

\begin{figure}
\center
\resizebox{\columnwidth}{!}{\includegraphics{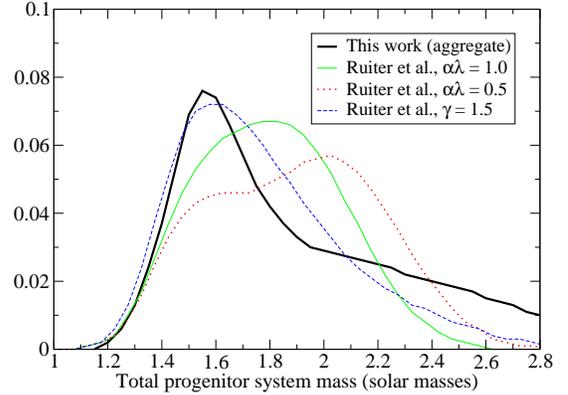}}
\caption{\small Total progenitor mass PDF for our sample
(thick solid line), compared with various population synthesis models of
\citet{ruiter09} as shown in \citet{fryer10}:
$\alpha\lambda = 1.0$ (thin solid line),
$\alpha\lambda = 0.5$ (dotted line), and
$\gamma = 1.5$ (dashed line).  Population synthesis models have been cut off
below 1.4~\Msol\ and convolved with the mass resolution of our reconstruction
technique, then normalized to match our total event rate.}
\label{fig:popmtot}
\end{figure}

Figure~\ref{fig:popmtot} presents the aggregate PDF of the masses of the SNe
in our sample, each weighted inversely by their control time in the SNfactory
search, accounting for both apparent magnitude and light-curve width.
Its lower edge is roughly consistent with the Chandrasekhar mass, with a tail
towards higher masses.

This distribution is qualitatively similar to expectations for
super-Chandrasekhar-mass merging white dwarf system masses from population
synthesis models \citep{ruiter09,fryer10}.  The limited statistics and
theoretical uncertainties make it difficult to draw a firm, quantitative
correspondence between observations and theory, but we can nevertheless
look for gross inconsistencies.  Figure~\ref{fig:popmtot} therefore
also includes three theoretical distributions of the progenitor system mass
(\Mtot) shown in \citet{fryer10}, cut off below 1.4~\Msol\ and convolved with
a two-sided Gaussian ($\sigma_- = 0.08\,\Mtot$, $\sigma_+ = 0.10\,\Mtot$)
to model our reconstruction precision.

Our observations correspond well with the \citet{ruiter09} models on the
leading edge, particularly the $\gamma = 1.5$ model.  The differences are
more pronounced on the high-mass tail; the agreement may improve with
a larger sample of SNe.  The other two models agree less well, predicting
more high-mass mergers than we see in our sample.
In the $\gamma = 1.5$ model, the relative rate of events $dN/d\Mtot$ is
about a factor of 3 higher at 1.4~\Msol\ than at 2.1~\Msol, consistent with
our observations.  The other two models predict more high-mass mergers than
we observe; we see no evidence for a separate formation channel or peak near
2~\Msol\ as suggested by \citet{fryer10}.  If such massive mergers do exist,
they must therefore produce less \nickel, and be less luminous at maximum
light, than the SNe in our sample.

\subsection{Hubble Residuals and Implications for Cosmology}
\label{subsec:hubble}

\begin{figure}
\center
\resizebox{\columnwidth}{!}
{\includegraphics{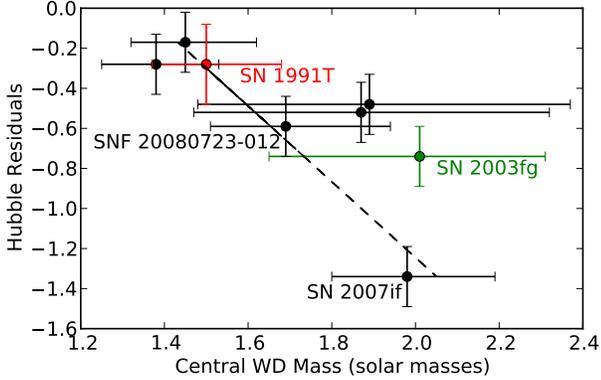}}
\caption{\small Hubble residuals vs. central merger remnant mass
for our sample (black points).  SN~1991T (red) and SN~2003fg (green) are also
plotted for comparison.  Error bars are 68\% ($1\sigma$) uncertainties.
The dotted black line is the best-fit linear trend.}
\label{fig:HvMWD}
\end{figure}

Table~\ref{tbl:salt2fit} also contains the Hubble residuals of each SN in our
sample from a standard \lcdm\ cosmology:
\begin{equation}
\Delta \mu = m_B - 25 - 5\, \log_{10} (d_L/\mathrm{Mpc}) 
                    - \alpha(s-1) + \beta c - M_B,
\end{equation}
where $m_B$, $s$ and $c$ are derived from the SALT2 fits,
$d_L$ is calculated using \citet{nedwright}, and we use values of
$\alpha$, $\beta$, and $M_B$ from the $w = -1$ fit of \citet{snls3yr-cosmo}.
(We use $M_B = M^1_B$, i.e., the absolute magnitude appropriate for SNe
in host galaxies with stellar mass less than $10^{10}~\Msol$.)

Figure~\ref{fig:HvMWD} plots the relation $\Delta \mu$ vs. \MWD\
(instead of \Mtot, so that it does not depend on interpreting the velocity
plateau in terms of an envelope)
for the six SNfactory SNe~Ia. For comparison we also plot SN~1991T and
SN~2003fg, modeled
in a similar fashion (see Appendix~\ref{sec:litsne}).
The Hubble residuals show a correlation with increasing mass
(Pearson $r = -0.76$, $p = 0.03$).  SN~2007if is 1.3~mag overluminous
for its stretch and color, and by 1~mag even if no correction is made for its
redder-than-average color.  SNF~20080723-012 is about 0.6~mag too luminous
for its stretch and color, putting it off the Hubble diagram but to a
lesser degree than SN~2007if.  At the low-mass end, SNF~20070803-005
and SNF~20080522-000, while slightly overluminous, are not alarmingly
discrepant, being within two standard deviations of the \lcdm\ Hubble diagram
assuming an intrinsic dispersion of 0.15~mag.
The higher-redshift ($z \sim 0.12$) SNF~20070528-003 and SNF~20070912-000
are too luminous for their stretch and color by about 0.5~mag; the residual
decreases to about 0.4~mag if no correction is made for the color.

If the correlation shown in Figure~\ref{fig:HvMWD} is real, and the large
Hubble residuals of SN~2003fg and SN~2007if are related to their large
progenitor masses, the large Hubble residuals of SNF~20070528-003 and
SNF~20070912-000 provide further circumstantial evidence that they too
may have super-Chandrasekhar-mass progenitors.  One might expect deviations
of super-Chandra candidate SNe~Ia from the width-luminosity relation for
normal SNe~Ia, either because the large ejected mass could increase the
diffusion time of radiation through the ejecta relative to Chandrasekhar-mass
explosions, or because the enhanced mixing of \nickel\ into the outer layers
of ejecta in these events affects the ionization state at the photosphere
\citep[see][and references therein]{kw07}.

Previous authors have warned about overluminous SNe~Ia that could skew
the cosmological parameters. \citet{reindl05} found that the SN~1991T
subclass give a mean Hubble residual of $-0.4$~mag. The sample presented
here has a mean Hubble residual is $-0.56$~mag, dropping to $-0.41$~mag
if SN~2007if were excluded on the grounds that it would be easy to eliminate.
However, without a spectroscopic veto at high redshift, or by relying only
on photometry at any redshift, it will be difficult to tag and remove
SNe such as these from the Hubble diagram.  Although SN~2007if is excessively
overluminous and would probably be excluded from any Hubble diagram on that
basis, somewhat less luminous analogues like SNF~20080723-012 probably could
not be safely removed based on their overluminosity alone.  It would also
be difficult, without spectroscopy, to distinguish intrinsically overluminous
SNe~Ia from normal SNe~Ia magnified by gravitational lensing; since lensing
conserves photons in general relativity \citep{weinberg76}, all events must
be included on the Hubble diagram to avoid bias in the reconstructed
cosmological parameters.  

Suppose candidate super-Chandra SNe~Ia occur at the relative rate we
calculate for our redshift-limited subsample in
\S\ref{subsec:rates}
($2.6^{+1.9}_{-0.7}$) and have a mean Hubble diagram residual of $-0.41$~mag,
the mean of the Table~\ref{tbl:salt2fit} entries excluding SN~2007if.
The mean absolute magnitude of SNe~Ia is then 0.01--0.02~mag higher (68\%~CL)
than it would be if these SNe followed the normal relations.  This number is
already comparable to the 2\% error budget suggested in \citet{kim04}
for Stage~IV SN~Ia cosmology experiments, and would be realized if the rate
evolves by either doubling or going to zero at high redshift.


\section{Conclusions}
\label{sec:conclusions}


We have observed a sample of SNe~Ia which we selected on the basis of their
pre-maximum similarity to SN~2007if, searching for new candidate
super-Chandra SNe~Ia.  Based on these spectroscopic analogues,
we find that SN~2007if probably lies on the extreme luminous end, 
with a hot, highly ionized photosphere
and a disturbed density structure resulting from a stalled shock within
the ejecta.  We interpret these characteristics as evidence for an
interaction with a circumstellar envelope early in the SN's evolution,
as one might expect in a double-degenerate merger scenario.

To reconstruct the masses of the progenitors, we apply an updated version
of the modeling technique of \citet{scalzo10}.
By applying this technique to SN~1991T for comparison with \citet{stritz06},
we show that including NIR flux near phase $+40$~days as part of the
bolometric flux has a significant impact on the final reconstructed mass;
while there is some uncertainty associated with correcting for NIR flux
not actually observed, ignoring such a correction causes the mass to be
systematically underestimated.  Our modeling technique includes a set of
Bayesian priors motivated by theoretical models, capturing the covariance
between different parameters of the system, and we sample the full
probability distribution using a Monte Carlo Markov chain.  From our sample,
SNF~20080723-012 now joins SN~2007if in having a super-Chandrasekhar-mass
progenitor at high statistical confidence.  The mean error-weighted system
mass for the four remaining
SNe in our sample ($1.63~\Msol \pm 0.10~\Msol$) is also above
the Chandrasekhar mass at a statistical confidence level exceeding 99\%.


Although the spectrophotometric observations discussed here are quite
detailed, with excellent temporal and wavelength coverage, the modeling
presented is relatively simple and is meant to explore a large parameter
space of rudimentary explosion models quickly and efficiently.
Detailed comparison
of the spectra to synthetic spectra of a range of contemporary explosion
models will be necessary to determine the extent to which our modeling
faithfully represents the density structure found in these SNe.  Our simple
model also assumes spherical symmetry of the SN ejecta, and this
approximation may not hold true in a significantly aspherical merger scenario.
The shell mass estimates must necessarily be subject to systematic
uncertainties of this sort, and hence might be more properly construed as
upper limits for which equality holds in the case of spherical symmetry.
The total mass estimates, which concern the bulk ejecta and not only the
outer layers, should be more robust.

Further study of a larger sample of such SNe~Ia
may provide additional insight into both the explosions themselves and the
formation of their progenitor systems, particularly the extent to which
velocity plateaus are linked to super-Chandrasekhar-mass progenitors.
Future data sets including not only optical-wavelength observations, but
infrared observations and nebular spectra, can help to verify the luminosity,
\nickel\ content and distribution, and total mass of these events.
Ultraviolet and X-ray observations of SNe~Ia spectroscopically similar to
SN~1991T at early phase may also help to
constrain the contribution of prompt shock emission to their luminosity,
and the density and spatial extent of any envelope of material that might
surround their progenitors just prior to explosion.  Polarimetry and
spectropolarimetry, as well as velocity offsets measured from nebular spectra,
can help constrain the degree of large-angular-scale asymmetry.
Finally, similar analysis of a much larger SN~Ia data set may help to
discover super-Chandrasekhar-mass SNe~Ia whose spectroscopic appearance
is unlike those in our sample; 
this may help to clarify the connection of these events to the general SN~Ia
population, with dividends paid to the understanding of SN~Ia progenitors
and to the precision of SN~Ia cosmology.




\acknowledgments

The authors are grateful to the technical and scientific staffs of the
University of Hawaii 2.2~m telescope, the W. M. Keck Observatory, Lick
Observatory, SOAR, and Palomar Observatory, to the QUEST-II collaboration,
and to HPWREN for their assistance in obtaining these data.
The authors wish to recognize and acknowledge the very significant cultural
role and reverence that the summit of Mauna Kea has always had within the
indigenous Hawaiian community.  We are most fortunate to have the opportunity
to conduct observations from this mountain.
This work was supported by the Director, Office of Science, Office of
High Energy Physics, of the U.S. Department of Energy under Contract No.
DE-AC02-05CH11231; by a grant from the Gordon \& Betty Moore Foundation;
and in France by support from CNRS/IN2P3, CNRS/INSU, and PNC.
RS acknowledges support from ARC Laureate Grant FL0992131.
YC acknowledges support from a Henri Chretien International Research
Grant administrated by the American Astronomical Society, and from the
France-Berkeley Fund.
This research used resources of the National Energy Research Scientific
Computing Center, which is supported by the Director, Office of Science,
Office of Advanced Scientific Computing Research, of the U.S. Department
of Energy under Contract No. DE-AC02-05CH11231.  We thank them for a generous
allocation of storage and computing time.
HPWREN is funded by National Science Foundation Grant Number ANI-0087344,
and the University of California, San Diego.
IRAF is distributed by the National Optical Astronomy Observatories,
which are operated by the Association of Universities for Research
in Astronomy, Inc., under cooperative agreement with the National
Science Foundation.  The spectra of SN~1991T were obtained through
the SUSPECT Supernova Spectrum Archive, an online database maintained
at the University of Oklahoma, Norman.
The distance modulus estimates for SN~1991T were obtained from
the NASA/IPAC Extragalactic Database (NED) which is operated by
the Jet Propulsion Laboratory, California Institute of Technology,
under contract with the National Aeronautics and Space Administration.
The SMARTS 1.3m observing queue receives support from NSF grant AST-0707627.
Part of this work was conducted at the Aspen Center for Physics during the
August 2010 workshop, ``Taking Supernova Cosmology into the Next Decade'',
and we gratefully acknowledge the ACP's hospitality in providing a
fruitful working environment.
We thank Dan Birchall for his assistance in collecting data with SNIFS.
RS also thanks Ashley Ruiter for providing data from population synthesis
models of double-degenerate mergers, and Brian Schmidt, Stuart Sim, and
Stefan Taubenberger for helpful discussions.

{\it Facilities:}
\facility{UH:2.2m (SNIFS)},
\facility{PO:1.2m (QUEST-II)},
\facility{CTIO:1.3m (ANDICAM)},
\facility{Keck:I (LRIS)},
\facility{Shane (Kast)},
\facility{SOAR (GHTS)}


\appendix

\section{Mass Modeling of Spectral Archetypes}
\label{sec:litsne}

In addition to SN~2007if, SN~2003fg and SN~1991T are spectroscopic
archetypes for the candidate super-Chandra SNe~Ia presented in this paper.
Thus, in order to have self-consistent mass estimates for the full archetype
set and provide the reader with familiar points for comparison, 
we have applied our modeling procedure to SN~199T and SN~2003fg.


\subsection{SN~1991T}


SN~1991T is the most well-known spectroscopic archetype of our sample,
although not the most spectroscopically extreme example.  SN~1991T has good
wavelength coverage in published spectra taken between maximum light and
phase $+40$~days, allowing us to use the same analysis techniques as for
the rest of our sample.  In addition, modeling of SN~1991T allows
a useful direct comparison to the earlier work of \citet{stritz06}.

SN~1991T's remarkably flat velocity evolution was pointed out by
\citet{phillips92}, and it has the second-lowest velocity gradient $\dot{v}$
in the original sample of \citet{benetti05}.
The DET2ENV2 model fits SN~1991T \citep{hk96},
and indeed SN~1991T was once suspected to have had a super-Chandrasekhar-mass
progenitor, mostly because of its high luminosity \citep{fisher99}.
Later measurements of the distance \citep{richtler01,gs01} to the
host galaxy, NGC~4527, have decreased the SN's luminosity considerably,
and hence its inferred mass.

We can make a new estimate of the progenitor mass of SN~1991T using our
reconstruction method, which considers the gamma-ray trapping in more detail
than previous studies.  We use the $BVRI$ light curve of \citet{lira98}
and the spectral time series of \citet{mdt95} obtained through the
SUSPECT online supernova spectrum archive.
\citet{kanbur03} give the Cepheid-based distance modulus to NGC~4527,
after correction for metallicity, as
$30.71 \pm 0.13$ (LMC period-luminosity relation) and
$30.78 \pm 0.13$ (Milky Way period-luminosity relation),
We adopt the mean of these two measurements ($30.74 \pm 0.13$) for our work.
From a joint fit to the spectral time series we extract
\ewna$ = 0.84 \pm 0.18$~\AA, giving $E(B-V)_\mathrm{host} = 0.13 \pm 0.03$
from the shallow TBC relation, in good agreement with the estimates of
\citet{phillips92} and \citet{phillips99}.
Our model represents a good fit to the available data, with
$\chi^2_\nu = 0.83$ ($P_\mathrm{fit} = 0.60$).
We recover a \nickel\ mass of $0.77 \pm 0.15$~\Msol\ and a white dwarf mass
$1.50^{+0.18}_{-0.13}$~\Msol.  If we suppose that the plateau is caused by
interaction with a uniform spherical carbon/oxygen envelope, this envelope
has a mass of about 0.1~\MWD, and the total system mass estimate goes up to
$1.65^{+0.22}_{-0.16}$~\Msol.  The results are similar to those for the
nominal Chandrasekhar-mass SNe in our sample, SNF~20070803-005
and SNF~20080522-000.

\citet{stritz06} measured a mass of $1.21 \pm 0.36$~\Msol\ for SN~1991T.
Our median derived mass for SN~1991T exceeds that of \citet{stritz06}
by over 0.4~\Msol; our 98\%~CL lower limit on \MWD\ is 1.28~\Msol.
However, we find that we can reproduce \citet{stritz06}'s numbers for $t_0$,
\nickel\ mass, and ejected mass if we neglect NIR corrections and ignore
covariances in our fitting procedure, particularly that between $q$ and $v_e$.
In particular, by including NIR corrections we find $t_0 = 48$~days,
much larger than the \citet{stritz06} value of 34~days and suggesting more
massive ejecta.  Our larger median value for $q$ (0.42) and slightly lower
$v_e$ (2850~\kms) pull in the other direction, resulting in a mass closer
to the the Chandrasekhar mass.

In summary, SN~1991T shows the observational hallmarks of a tamped detonation
and may well have been a double-degenerate merger.  Because of uncertainties
in the host distance and reddening correction, we are unable to establish
with confidence whether SN~1991T was itself super-Chandrasekhar-mass.


\subsection{SN~2003fg}

SN~2003fg \citep{howell06} has a comparatively limited data set for our
purposes, with only one spectrum at maximum light and no photometry at
sufficiently late times for us to constrain the gamma-ray transparency.
Nevertheless, there are noted similarities to SN~2007if:
The single spectrum available is a good match to SN~2007if at a comparable
phase, though with stronger \ion{Si}{2} and \ion{S}{2}.
The observed $i'$-band light curve of SN~2003fg shows no significant
inflection, like SN~2007if and unlike the less massive SNfactory SNe~Ia.
The low ($\sim8000~\kms$) \ion{Si}{2} velocity near maximum light is also
like SN~2007if at similar phases, and makes SN~2003fg a good candidate
for a tamped detonation, although SN~2009dc showed a similarly low velocity
without a discernable plateau \citep{yamanaka09}.

We can obtain a lower limit on the mass of SN~2003fg assuming it can be
fit by a tamped detonation.  We use the value of the bolometric absolute
magnitude at maximum light ($M_\mathrm{bol} = -19.87$)
and the velocity of the \ion{Si}{2}~$\lambda 6355$ absorption minimum
($\sim8000\pm500~\kms$) derived by \citet{howell06}.
Assuming no host galaxy reddening, we find a \nickel\ mass of
$1.18 \pm 0.16~\Msol$ and a 98\% CL lower limit on the total system mass
of $1.77 \Msol$ ($1.46~\Msol$ if the envelope mass is neglected).
These are somewhat looser constraints than those of \cite{howell06},
but are free of the assumption that the photospheric velocity should be
related directly to the kinetic energy released in the explosion, which,
as we have seen, may not be a good approximation for some density structures.


\end{document}